\documentclass[aps,prx,superscriptaddress,amsmath,amssymb,twocolumn,showpacs,floatfix,reprint]{revtex4-2}
\usepackage{graphicx}
\usepackage{dcolumn}
\usepackage{cancel}
\usepackage{bm}
\usepackage{tikz}
\usepackage{siunitx}
\usepackage{tablefootnote}
\usepackage{physics}
\usepackage[section]{placeins}
\definecolor{darkblue}{rgb}{0,0,.65}
\definecolor{darkgreen}{rgb}{1,0,0}
\usepackage{nicefrac}
\usepackage{booktabs}
\usepackage[colorlinks=true, urlcolor=blue, linkcolor=blue, citecolor=blue, pdftex]{hyperref}

\newcommand{\s}{\bm{s}}

\usepackage[english]{babel}
\usepackage{bm}
\usepackage{amssymb}
\usepackage{color, soul}
\usepackage{amsmath}
\usepackage{amstext}
\usepackage{latexsym}
\usepackage{graphicx}
\usepackage{mathtools}
\usepackage[ruled,vlined]{algorithm2e}

\begin{document}
\title{Tensor Network Markov Chain Monte Carlo: Efficient Sampling of Three-Dimensional Spin Glasses and Beyond}

\author{Tao Chen}
\affiliation{
Hefei National Laboratory for Physical Sciences at the Microscale and Department of Modern Physics, University of Science and Technology of China, Hefei 230026, China}
\affiliation{Hefei National Laboratory, University of Science and Technology of China, Hefei 230088, China}

\author{Jing Liu}
\email{jing.liu@itp.ac.cn}
\affiliation{Institute of Theoretical Physics, Chinese Academy of Sciences, Beijing 100190, China}

\author{Youjin Deng}%
\email{yjdeng@ustc.edu.cn}
\affiliation{
Hefei National Laboratory for Physical Sciences at the Microscale and Department of Modern Physics, University of Science and Technology of China, Hefei 230026, China}
\affiliation{Hefei National Laboratory, University of Science and Technology of China, Hefei 230088, China}

\author{Pan Zhang}%
\email{panzhang@itp.ac.cn}
\affiliation{Institute of Theoretical Physics, Chinese Academy of Sciences, Beijing 100190, China}
\affiliation{
School of Fundamental Physics and Mathematical Sciences, Hangzhou Institute for Advanced Study, UCAS, Hangzhou 310024, China
}

\date{\today}

\begin{abstract}
Sampling the three-dimensional (3D) spin glass—i.e., generating equilibrium configurations of a 3D lattice with quenched random couplings—is widely regarded as one of the central and long-standing open problems in statistical physics. The rugged energy landscape, pronounced critical slowing down, and intrinsic ergodicity breaking render standard Monte Carlo methods severely inefficient, particularly for large systems at low temperatures. In this work, we introduce the Tensor Network Markov Chain Monte Carlo (TNMCMC) approach to address the issue. It generates large-scale collective updates in MCMC using tensor networks on the 2D slices of the 3D lattice, greatly improving the autocorrelation time and offering orders-of-magnitude speed-ups over conventional MCMC in generating unbiased samples of the Boltzmann distribution. We conduct numerical experiments on 3D spin glasses up to system size $64\times 64\times 64$ using a single CPU, and show that TNMCMC dramatically suppresses critical slowing down in large disordered systems, which usually require a supercomputer to perform MCMC simulations. Furthermore, 
we apply our approach to the 3-state Potts model up to system size $64\times 64\times 64$ using a single CPU, and show that the TNMCMC approach efficiently traverses the exponential barriers of the strong first-order transition, whereas conventional MCMC fails. Our results reveal that TNMCMC opens a promising path toward tackling long-standing, formidable three-dimensional problems in statistical physics.

\end{abstract}

\maketitle

\noindent
Three-dimensional spin glasses combine the lattice geometry, quenched disorders, and frustrations, exhibiting a rugged energy landscape~\cite{edwards1975theory,binder1986spin,young2006numerical}. They are pivotal in statistical physics for providing profound insights into the nature of phase transitions~\cite{nishimori2010elements}, testing competing theories of replica symmetry breaking, benchmarking sampling and optimization algorithms, and providing the baseline for understanding glassy dynamics and aging phenomena.
However, they remain one of the last unsolved challenges in statistical mechanics. Yet, their equilibrium and dynamical properties remain analytically intractable, despite half a century of effort. 

For decades, Markov-chain Monte Carlo (MCMC)~\cite{newman1999monte,landau2014guide} (most commonly single-spin-flip Metropolis algorithms~\cite{metropolis1953equation}) has been the standard workhorse for sampling and studying three-dimensional (3D) spin glasses. In this paradigm, one constructs a Markov chain whose stationary distribution is the Boltzmann distribution at a given temperature; successive configurations are generated by proposing local spin flips and accepting them according to the Metropolis–Hastings scheme.  
However, the standard single-spin-flip MCMC is severely inefficient at a low temperature due to the critical slowing down caused by the rugged energy landscape. The correlation time grows quickly with the system size and inverse temperature $\beta$, and the spin flips become locked in exponentially numerous metastable valleys, so the Markov chain needs exponentially many flips to reach equilibrium. 
Although advanced techniques such as parallel tempering and population annealing can accelerate the mixing of the Markov chain at low temperatures, it is still difficult to study large systems due to the rapid growth of the computational cost with system size. 

An effective way to address the critical slowing down of MCMC is the \textit{cluster update}, which updates a cluster (or block) of spins, creating a configuration that is much less correlated with the previous configuration than the case of single-spin-flip. 
The examples include the celebrated Swendsen-Wang 
and Wolff cluster algorithms~\cite{swendsen1987nonuniversal,wolff1989collective},
which can dramatically enhance efficiency for ferromagnetic systems, 
due to their intrinsic connection to the Fortuin-Kasteleyn representation~\cite{fortuin1972random-cluster}.
However, it remains an open question to develop a cluster method that can correctly reflect spin-spin correlations in spin glass systems, even though the alleviation of critical slowing-down in 3D~\cite{zhu2015efficient} has been reported by using a variant of Houdayer's algorithm~\cite{houdayer2001cluster}. 


Here, we introduce the Tensor Network Markov Chain Monte Carlo (TNMCMC) to tackle the sampling challenge in 3D spin glasses.
By mapping 2D slices of the 3D lattice into 2D tensor networks, we generate whole-slice spin configurations via tensor-network contraction, delivering large-scale collective updates that leap over local minima and suppress the critical slowing-down. 
The bias of generated samples induced in 2D tensor network contractions is 
eliminated through a Metropolis-Hastings algorithm, ensuring detailed balance.
To benchmark the efficacy of our approach, we apply it to the 3D Edwards-Anderson (EA) spin glass up to size $64\times 64\times 64$. The results show that TNMCMC dramatically suppresses critical slowing down at low temperatures, reaching equilibrium many orders faster than the standard MCMC in both the number of MC steps and in computational time. The advantage of our approach increases quickly as the system size increases and the temperature decreases. We further apply TNMCMC to the 3-state Potts model, which exhibits a strong first-order phase transition that is challenging for standard MCMC, and we show that TNMCMC can overcome the exponential tunneling barrier efficiently. In the following text, we will first introduce the TNCMCM approach, then present the numerical results in 3D spin glasses and the Potts model in detail, and discuss the further generalizations of TNMCMC.

\paragraph*{TNMCMC---}
Consider the Boltzmann distribution of $n$ spins, 
$$P(\sigma) = \frac{1}{Z}e^{-\beta E(\mathbf s)},$$
where ${\mathbf s} \in\{+1,-1\}^n$ is a configuration, $\beta$ is the inverse temperature, $E(\s)$ is the energy function, and $Z=\sum_{\mathbf s}e^{-\beta E(\mathbf s)}$ denotes the partition function. Sampling of the Boltzmann distribution belongs to the computational class of \#P problems, and there do not exist exact algorithms for solving the general problem in polynomial time.
In standard MCMC, a Markov chain of configurations is constructed for generating samples. At each step, given the current state $\s$, a candidate $\s^\prime$ is drawn from a proposal distribution $q(\s^\prime)$ and accepted with probability
\begin{equation}\label{eq:acceptance}
    A(\s \rightarrow \s^\prime) = \min \left\{1,  \frac{q(\s) \times e^{-\beta E(\s^{\prime})}}{q(\s^{\prime}) \times e^{-\beta E(\s)}}\right\}.
\end{equation}
The Metropolis algorithm~\cite{metropolis1953equation} uses a candidate $\s^{\prime}$ that differs from $\s$ with only one spin. 
This method has a rather high acceptance probability but cannot cross a large energy barrier, 
and thus it may need exponentially many steps to decorrelate if there is one and more free energy barriers.

\begin{figure}[!t]
  \includegraphics[width = 0.9\linewidth]{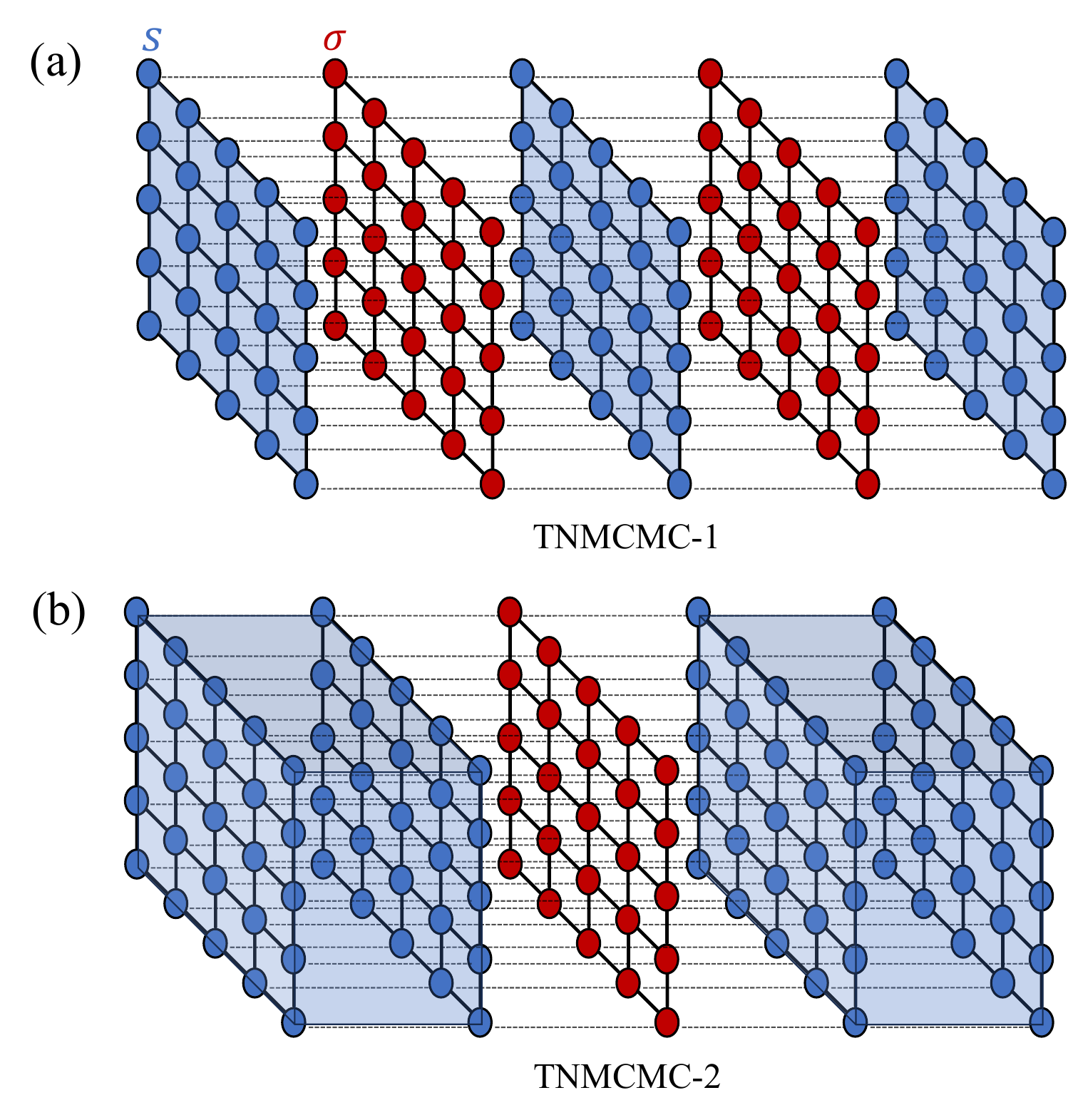}
  \caption{\textbf{Schematic of the Tensor-Network Markov-Chain Monte Carlo (TNMCMC) for efficient sampling of 3D spin-glasses.} At each step of sampling, spins are separated into two groups: sampling spins with configuration $s$ (in the blue-shaded slices of the 3D lattice) and fixed spins with configuration $\sigma$ (red spins).
   We present two sampling-spin schemes: (a) the spins to be sampled lie on alternating 2D slices across the 3D lattice, separated by the fixed spins; (b) the sampling spins form pairs of two adjacent slices which can be merged into a single ``superspin'' layer.  
For each scheme, the conditional probability $P(s|\sigma)$ is computed approximately with a tensor network, and configurations $s$ are drawn and then resampled via the Metropolis–Hastings rule (Eq.\eqref{eq:acceptance}) to remove the bias introduced by the approximate tensor network contractions.
  }
  \label{fig:tnmcmctensor}
\end{figure}

To address this issue, we propose to update an entire correlated cluster of spins filling slices of the 3D lattices using tensor networks. We describe our approach using an illustrative example with size $5\times 5\times 5$ as shown in Fig.~\ref{fig:tnmcmctensor}. At each step, the spins are (dynamically) divided into two groups, the sampling spins with configuration $\s$ and fixed spins with configuration $\sigma$. In the figure, we illustrate two different divisions explored in this work: TNMCMC-1 splits the 3D lattice into alternating layers with sampling spins occupying every odd layer, fixed spins every even layer (and vice versa); TNMCMC-2 assigns sampling spins to two contiguous layers, leaving the middle layer fixed. During each Markov step, the full spin configuration is updated from $(\s, \sigma)$ to $(\s^{\prime}, \sigma)$: We first approximately compute the conditional probability $P(\s^\prime | \sigma)$ using tensor networks~\cite{schollwöck2011density-matrix,orús2014practical,ran2020tensor,cirac2021matrix,pan2022simulation}, producing a proposal distribution $q(\s^\prime|\sigma)$ and drawing candidate configurations~\cite{chen2025tensor} $\s^\prime$ from it. These candidates are then accepted or rejected via the Metropolis rule (Eq.\eqref{eq:acceptance}) to produce a final $\s^\prime$ which is set as a current configuration for the next Markov step.

\begin{figure*}[!t]
  \includegraphics[width=\linewidth]{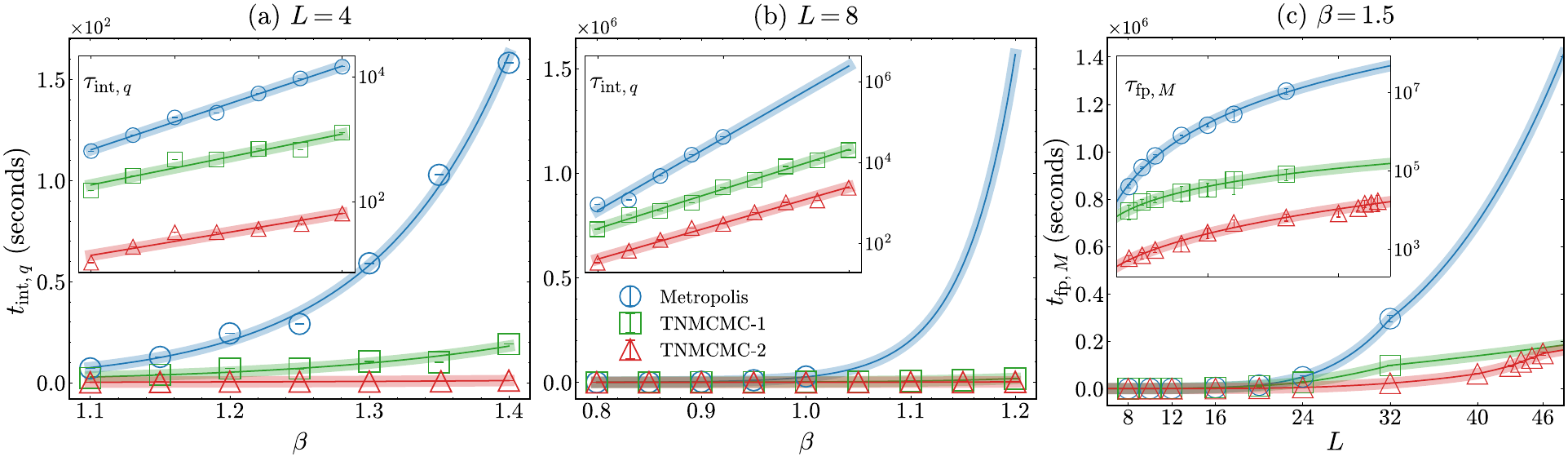}
  \caption{\textbf{Results for the 3D Edwards-Anderson spin glasses.} 
  (a,b) Integrated autocorrelation time of the spin-glass overlap parameter, $\tau_{\text{int},q}$, and the computational cost per independent sample, $t_{\text{int},q}$, for Tensor Network Markov Chain Monte Carlo (TNMCMC) and conventional Metropolis algorithm on system sizes $L=4$ and $L=8$.
  The computational cost is given by $t_{\text{int},q}=\tau_{\text{int},q} \times t_s$, with $t_s$ representing the wall-clock time per sweep on a single CPU. 
  (c) System-size dependence of the first-passage time to zero magnetization, $\tau_{\text{fp}, M}$, along with the wall-clock time, $t_{\text{fp},M}=\tau_{\text{fp}, M} \times t_s$, at $\beta=1.5$.
  In the figures, each data point is averaged over $10$ disorder instances, and the solid line is a fit with $A\exp(B*L^C)$, where $A, B$, and $C$ are fitting parameters. In the figures, the data is averaged across disorders, with the number of disorders ranging from 12 to 1024.}
  \label{fig:spinglass}
\end{figure*}

In detail, the joint distribution of sampling spins $\s$ is factorized into a product of conditional probabilities $$P(\s|\sigma) = \prod_iP(s_i|\s_{<i},\sigma),$$ where $\s_{<i}$ denotes the sampling spins in front of spin $i$. Each conditional probability \begin{equation}P(s_i|\s_{<i},\sigma)=Z(s_i,\s_{<i},\sigma)/\sum_{s_i}Z(s_i,\s_{<i},\sigma),\end{equation} 
where $Z(s_i,\s_{<i},\sigma)=\sum_{\s_{>i}}e^{-\beta E(s_i,\s_{<i},\s_{>i},\sigma)}$ can be computed using tensor network contractions~\cite{pan2020contracting,liu2021tropical} to sum over $\s_{>i}$, configurations of sampling spins after $i$. Then each sampling spin is sampled one by one using the corresponding conditional probability, resulting in $\s$.

For small systems as shown in Fig.~\ref{fig:tnmcmctensor}, sampling of $\s$ can be made exact using tensor networks. 
Whenever $Z(s_i,\s_{<i},\sigma)$ and $P(s_i|\s_{<i},\sigma)$ are computed exactly, the acceptance probability becomes unity, because the protocol realizes the exact sampling. However, exact sampling of collective spins belongs to the computational class of \#P, and for larger systems with a larger cluster of sampling spins, $Z(s_i,\s_{<i},\sigma)$ has to be computed approximately, which will lower the acceptance probability Eq.~\eqref{eq:acceptance}. Both the shape and size of the cluster affect contraction error and acceptance probability, forcing a trade-off between the cluster size and acceptance probability. A single-spin flip yields a minimal cluster containing one spin and a high acceptance, whereas a cluster spanning the entire 3D lattice incurs prohibitive contraction costs and large truncation errors, driving the acceptance probability to zero (Supporting Materials). In TNMCMC, we choose 2D layer composed of spins (TNMCMC-1) or superspins (TNMCMC-2) to optimize this size-acceptance trade-off, based on the very efficient contraction methods designed for 2D tensor networks~\cite{okunishi2022developments,white1992density,nishino1998density,chen2011partial,xie2009second,fliu2022variational,orús2014practical,ran2020tensor,cirac2021matrix,xiang2023density,schollwöck2011density-matrix,chen2025tensor}. 
In detail, at each Markov step, we randomly select disjoint layers of spins (or super-spins), compute $Z(s_i,\mathbf{s}_{<i},\sigma)$ and the conditional probabilities via boundary-MPS~\cite{white1992density,orús2014practical,chen2025tensor} method with fixed bond dimension, draw a batch of candidate samples from these probabilities, and finally accept one via Metropolis rule, This yields the TNMCMC algorithm.

\begin{figure*}[!t]
  \includegraphics[width=\linewidth]{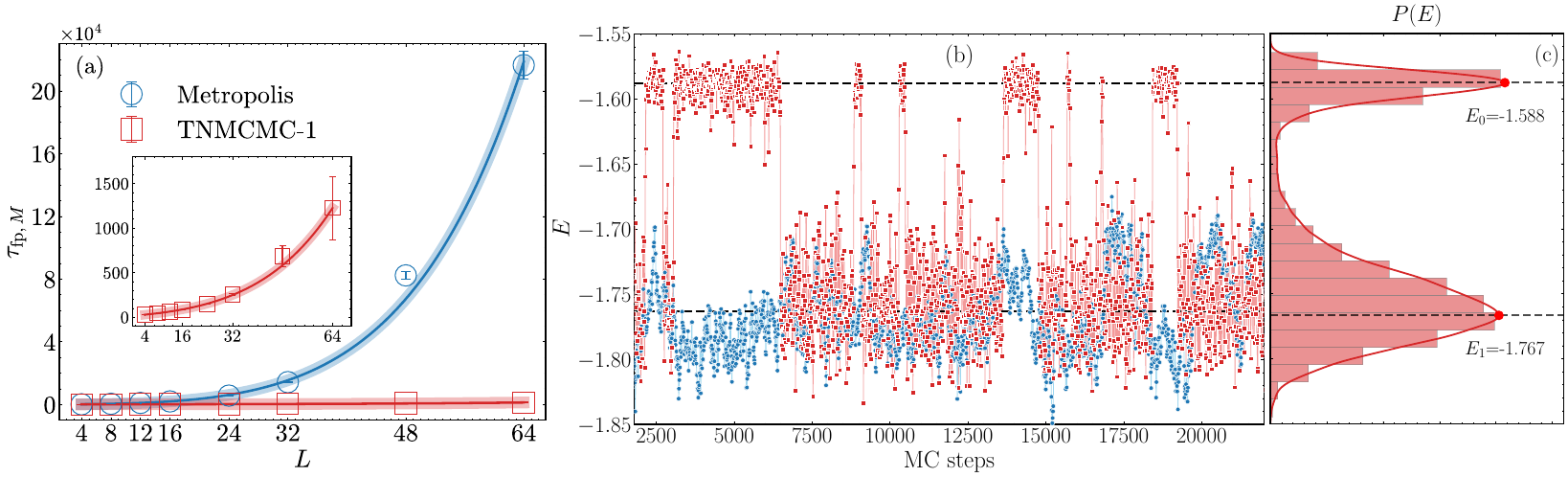}
  \caption{\textbf{Results for the 3D 3-state Potts model with first-order transition.} 
  (a) The first-passage time to zero magnetization, $\tau_{\text{fp}, M}$ obtained by the conventional Metropolis algorithm and TNMCMC-1 (with bond dimension $\chi=3$), at the critical $\beta_c = 0.550565$. Each point is the statistical average, with the number of instances ranging from 15 to 1024.
  (b) Energy of samples obtained using TNMCMC-1 (red) and Metropolis (blue) for a lattice with $L=48$ at $\beta_c$, where the system exhibits two energy barriers as illustrated by the histogram of TNMCMC-1 configurations in (c). Notice that while the conventional Metropolis traps into one energy valley, TNMCMC jumps back and forth between two energy barriers, efficiently traversing the exponential barriers of the strong first-order transition. 
  }
  \label{fig:firstorder}
\end{figure*}

\paragraph*{Sampling 3D spin glasses---}

To test our approach, we conduct numerical experiments for a prototype of spin glass, the 3D EA  model~\cite{edwards1975theory}, on an $L \times L \times L$ cubic lattice with free boundary conditions, with energy function
\begin{equation}
    E(\s) = -\sum_{\langle ij\rangle} J_{ij}s_i s_j,
\end{equation}
where $s_i \in \{-1,+1\}$, and the coupling $J_{ij}$ takes $1$ with probability $p$ and $-1$ with probability $1-p$.
For $p=0.5$, the system exhibits a spin-glass transition at temperature $T_c = 1.1019(29)$~\cite{baity-jesi2013critical}.
Below $T_c$, the system enters a glassy phase characterized by a rugged energy landscape and slow dynamics, and the standard MCMC methods are hampered by relaxation times that diverge exponentially with system volume size and inverse temperature.

To provide a quantitative assessment of the sampling efficiency, we compute the integrated autocorrelation time, $\tau_{\text{int},q}$, associated with the spin-glass overlap order parameter, $q=(1/L^3) \sum_{i} s_i^{(1)} s_i^{(2)}$, where $s_i^{(1)}$ and $s_i^{(2)}$ denote spin configurations from two independent replicas of the system.
For large-scale systems, the computational expense of evaluating $\tau_{\text{int},q}$ is substantial, and we introduce a complementary metric, the first‑passage time to zero magnetization $\tau_{\text{fp},M}$ from an initial configuration of all up spins. It measures the characteristic timescale required for the system to evolve from a fully magnetized, non-equilibrium state ($M=1$) to a state of zero magnetization for the first time, serving as a proxy for the escape time from metastable states within the system's complex energy landscape.

The results on the 3D spin glasses are shown in Fig.~\ref {fig:spinglass}. Fig.~\ref{fig:spinglass}(a,b) illustrates the temperature dependence of the integrated autocorrelation time, $\tau_{\text{int},q}$, for small systems with sizes $L = 4$ and $8$.
As expected, the Metropolis algorithm exhibits a pronounced increase in $\tau_{\text{int},q}$ at lower temperatures, a manifestation of critical slowing down, and its inability 
to efficiently explore the complex energy landscape.
In contrast, TNMCMC methods demonstrate a significant suppression of integrated autocorrelation times, especially within the glassy phase, and the gap is increasing quickly as $\beta$ increases.
We note that compared with the Metropolis method, there is an overhead of computational cost per TNMCMC sweep $t_s$ due to the tensor network contractions. However, the benefit of the cluster update in TNMCMC outweighs this overhead, leading to a substantial enhancement in overall efficiency. In the figure, we present the computational time required to generate a statistically independent sample, $t_{\text{int},q}=\tau_{\text{int},q} \times t_s$, where we can see that the overall computational time of TNMCMC is orders of magnitude smaller than that of Metropolis.
In Fig.~\ref{fig:spinglass}(c), we fixed an inverse temperature at $\beta=1.5$,
which is significantly below the critical temperature $\beta_c\approx 0.9$, and present the first-passage time to zero magnetization, $\tau_{\text{fp},M}$, as a function of system size. 
We also report the the observed exponential scaling of $\tau_{\text{fp},M}$ with system size,
\begin{equation}
  \tau_{\text{fp},M} \approx \tau_0 \exp(cL^\psi),
  \label{eq:taufit}
\end{equation}
to characterise the activated dynamics in glassy systems. A fit of the data to this form yields scaling exponents of $\psi \approx 0.56$ for Metropolis, $\psi \approx 0.16$ for TNMCMC-1, and $\psi \approx 0.0051$ for TNMCMC-2.
We have also performed experiments at other temperatures and observed qualitatively similar results. For example, at $\beta=1.0$, a temperature slightly lower than 
$\beta_c \approx 0.9$,
TNMCMC can successfully obtain results with system size up to $L=64$ with $\tau_{\text{fp},M}(\textrm{TNMCMC2})\approx 3.99\times 10^3$, while $\tau_{\text{fp},M}(\textrm{Metropolis})\approx 2.04\times 10^6$.

\paragraph*{Sampling the 3-state 3D Potts model---}
The 2D-slice update via tensor networks is an approach general to 3D systems. Here, we give an example using the 3-state Potts model in 3D, with the energy function defined as
\begin{equation}
  E(\sigma) = -J\sum_{\langle i,j\rangle}\delta(\sigma_i,\sigma_j),
\end{equation}
where $\sigma_i \in \{1,2,3\}$, $J>0$ is the ferromagnetic coupling strength, and $\delta()$ is the Kronecker delta function.
This model is known to exhibit a sharp first-order phase transition~\cite{janke1997three-dimensional}, where conventional MCMC algorithms typically struggle.
We conducted numerical experiments on the Potts model up to system size $64\times 64\times 64$ and present the scaling of the first-passage time to zero magnetization $\tau_{\text{fp},M}$ at its critical inverse temperature $\beta_c=0.550565$~\cite{janke1997three-dimensional}.
In this case, we employ periodic boundary conditions and use TNMCMC-1 only. At each step, a random slice is picked up, which can be on $xy$, $yz$ 
or $zx$ planes, and then, the spins on a randomly chosen line are fixed.
From Fig.~\ref{fig:firstorder}(a), we can see that the Metropolis algorithm exhibits a dramatic exponential growth of $\tau_{\text{fp},M}$ with increasing $L$, due to the first-order transition in the system.
This scaling behavior underscores the Metropolis algorithm's inefficiency in overcoming the large free-energy barrier that the system must overcome to transition from one phase to the other.
In stark contrast, TNMCMC-1 displays a significantly slower growth of $\tau_{\text{fp},M}$ on system size, yielding a much smaller autocorrelation time.
To further illustrate the dynamics of the simulation, Fig.~\ref{fig:firstorder}(b) displays the energy of a series of configurations generated by both the Metropolis and TNMCMC-1 algorithms.
We can see that the Metropolis dynamics are characterized by prolonged periods of trapping within a single phase.
In this specific simulation, it is trapped in an ordered state of 
dominant spins, and the energy fluctuation reflects the creation and annihilation dynamics of small-scale domains of other spins.
In contrast, the TNMCMC-1 simulation exhibits rapid and frequent transitions between the two energy levels associated with the ordered and disordered phases, 
and, further, effectively generates intermediate states from  
their coexistence,
as illustrated in the energy histogram in Fig~\ref{fig:firstorder}(c).
The results demonstrate the TNMCMC's capacity to explore the whole configuration space comprehensively, also indicating that the non-local, tensor-network-guided updates in TNMCMC facilitate more effective traversal of the free energy landscape even in 
the presence of a substantial free energy barrier.

\paragraph*{Conclusions and discussions---}
We have introduced TNMCMC, a general-purpose algorithm to simulate challenging 3D statistical physics systems efficiently. TNMCMC integrates tensor-network algorithms with MCMC sampling. By slicing the 3D lattice into 2D layers and sampling spins in each layer with tensor networks, the method delivers non-local, collective moves that leap across rugged energy landscapes. A final Metropolis–Hastings acceptance step rigorously enforces detailed balance, ensuring both ergodicity and correctness, while sidestepping the infeasible cost and large error of fully contracting 3D tensor networks.
We have demonstrated its efficacy on the 3D Edwards–Anderson Ising spin glass and the 3-state 3D Potts model, showing that TNMCMC significantly suppresses critical slowing down in the glassy phase and overcomes the free-energy barriers of a strong first-order phase transition.

Thanks to the general 2D tensor network contraction and sampling, TNMCMC is not confined to a specific system; it offers a general framework for tackling a wide class of 3D systems that are challenging to conventional MCMC methods. We also anticipate that TNMCMC will find applications in two-dimensional quantum systems, which can be mapped to 3D classical systems~\cite{liu2021accurate,liu2025accurate,layden2023quantumenhanced}. 

The CPU-based Julia implementation of TNMCMC is available at~\cite{tnmcmc-repo}. While the algorithm already performs strongly in 3D systems, the underlying two-dimensional tensor-network contractions and sampling implementation remain sub-optimal. 
we anticipate significant speed-ups through further engineering and GPU acceleration~\cite{weigel2012performance,bisson2025massivescale,bisson2025universal,chen2025batchtnmc}.
In the current version, the tensor network contractions are
performed frequently, since the to-be-updated slices are subject to effective external fields from their adjacent layers of fixed spins.  One might  generate (and dynamically update) a long list of ``typical"  external fields, so that tensor network contractions are only needed occasionally. 
This would introduce further approximations and reduce acceptance probabilities, requesting some compromise. 
While beyond the scope of the current work, a large-scale application of TNMCMC is promising in the future.

\begin{acknowledgments}
This work is supported by Projects 12325501, 12247104, 12275263, and 12405047 of the National Natural Science Foundation of China, and the Innovation Program for Quantum Science and Technology (under Grant No. 2021ZD0301900). YD thanks the support from the Natural Science Foundation of Fujian Province of China (under Grant No. 2023J02032).
\end{acknowledgments}


\bibliography{references.bib}

\begin{thebibliography}{37}%
\makeatletter
\providecommand \@ifxundefined [1]{%
 \@ifx{#1\undefined}
}%
\providecommand \@ifnum [1]{%
 \ifnum #1\expandafter \@firstoftwo
 \else \expandafter \@secondoftwo
 \fi
}%
\providecommand \@ifx [1]{%
 \ifx #1\expandafter \@firstoftwo
 \else \expandafter \@secondoftwo
 \fi
}%
\providecommand \natexlab [1]{#1}%
\providecommand \enquote  [1]{``#1''}%
\providecommand \bibnamefont  [1]{#1}%
\providecommand \bibfnamefont [1]{#1}%
\providecommand \citenamefont [1]{#1}%
\providecommand \href@noop [0]{\@secondoftwo}%
\providecommand \href [0]{\begingroup \@sanitize@url \@href}%
\providecommand \@href[1]{\@@startlink{#1}\@@href}%
\providecommand \@@href[1]{\endgroup#1\@@endlink}%
\providecommand \@sanitize@url [0]{\catcode `\\12\catcode `\$12\catcode
  `\&12\catcode `\#12\catcode `\^12\catcode `\_12\catcode `\%12\relax}%
\providecommand \@@startlink[1]{}%
\providecommand \@@endlink[0]{}%
\providecommand \url  [0]{\begingroup\@sanitize@url \@url }%
\providecommand \@url [1]{\endgroup\@href {#1}{\urlprefix }}%
\providecommand \urlprefix  [0]{URL }%
\providecommand \Eprint [0]{\href }%
\providecommand \doibase [0]{https://doi.org/}%
\providecommand \selectlanguage [0]{\@gobble}%
\providecommand \bibinfo  [0]{\@secondoftwo}%
\providecommand \bibfield  [0]{\@secondoftwo}%
\providecommand \translation [1]{[#1]}%
\providecommand \BibitemOpen [0]{}%
\providecommand \bibitemStop [0]{}%
\providecommand \bibitemNoStop [0]{.\EOS\space}%
\providecommand \EOS [0]{\spacefactor3000\relax}%
\providecommand \BibitemShut  [1]{\csname bibitem#1\endcsname}%
\let\auto@bib@innerbib\@empty
\bibitem [{\citenamefont {Edwards}\ and\ \citenamefont
  {Anderson}(1975)}]{edwards1975theory}%
  \BibitemOpen
  \bibfield  {author} {\bibinfo {author} {\bibfnamefont {S.~F.}\ \bibnamefont
  {Edwards}}\ and\ \bibinfo {author} {\bibfnamefont {P.~W.}\ \bibnamefont
  {Anderson}},\ }\bibfield  {title} {\bibinfo {title} {{Theory of spin
  glasses}},\ }\href {https://doi.org/10.1088/0305-4608/5/5/017} {\bibfield
  {journal} {\bibinfo  {journal} {J. Phys. F: Met. Phys.}\ }\textbf {\bibinfo
  {volume} {5}},\ \bibinfo {pages} {965} (\bibinfo {year} {1975})}\BibitemShut
  {NoStop}%
\bibitem [{\citenamefont {Binder}\ and\ \citenamefont
  {Young}(1986)}]{binder1986spin}%
  \BibitemOpen
  \bibfield  {author} {\bibinfo {author} {\bibfnamefont {K.}~\bibnamefont
  {Binder}}\ and\ \bibinfo {author} {\bibfnamefont {A.~P.}\ \bibnamefont
  {Young}},\ }\bibfield  {title} {\bibinfo {title} {{Spin glasses: Experimental
  facts, theoretical concepts, and open questions}},\ }\href
  {https://doi.org/10.1103/RevModPhys.58.801} {\bibfield  {journal} {\bibinfo
  {journal} {Rev. Mod. Phys.}\ }\textbf {\bibinfo {volume} {58}},\ \bibinfo
  {pages} {801} (\bibinfo {year} {1986})}\BibitemShut {NoStop}%
\bibitem [{\citenamefont {Young}(2006)}]{young2006numerical}%
  \BibitemOpen
  \bibfield  {author} {\bibinfo {author} {\bibfnamefont {A.}~\bibnamefont
  {Young}},\ }\bibinfo {title} {{Numerical {Simulations} of {Spin} {Glasses}:
  {Methods} and {Some} {Recent} {Results}}},\ in\ \href
  {https://doi.org/10.1007/3-540-35284-8_2} {\emph {\bibinfo {booktitle}
  {Computer Simulations in Condensed Matter Systems: From Materials to Chemical
  Biology Volume 2}}},\ \bibinfo {editor} {edited by\ \bibinfo {editor}
  {\bibfnamefont {M.}~\bibnamefont {Ferrario}}, \bibinfo {editor}
  {\bibfnamefont {G.}~\bibnamefont {Ciccotti}},\ and\ \bibinfo {editor}
  {\bibfnamefont {K.}~\bibnamefont {Binder}}}\ (\bibinfo  {publisher} {Springer
  Berlin Heidelberg},\ \bibinfo {address} {Berlin, Heidelberg},\ \bibinfo
  {year} {2006})\ pp.\ \bibinfo {pages} {31--44}\BibitemShut {NoStop}%
\bibitem [{\citenamefont {Nishimori}\ and\ \citenamefont
  {Ortiz}(2010)}]{nishimori2010elements}%
  \BibitemOpen
  \bibfield  {author} {\bibinfo {author} {\bibfnamefont {H.}~\bibnamefont
  {Nishimori}}\ and\ \bibinfo {author} {\bibfnamefont {G.}~\bibnamefont
  {Ortiz}},\ }\href {https://doi.org/10.1093/acprof:oso/9780199577224.001.0001}
  {\emph {\bibinfo {title} {Elements of Phase Transitions and Critical
  Phenomena}}}\ (\bibinfo  {publisher} {Oxford University Press},\ \bibinfo
  {year} {2010})\BibitemShut {NoStop}%
\bibitem [{\citenamefont {Newman}\ and\ \citenamefont
  {Barkema}(1999)}]{newman1999monte}%
  \BibitemOpen
  \bibfield  {author} {\bibinfo {author} {\bibfnamefont {M.}~\bibnamefont
  {Newman}}\ and\ \bibinfo {author} {\bibfnamefont {G.}~\bibnamefont
  {Barkema}},\ }\href {https://books.google.co.jp/books?id=KKL2nQEACAAJ} {\emph
  {\bibinfo {title} {Monte Carlo Methods in Statistical Physics}}}\ (\bibinfo
  {publisher} {Clarendon Press},\ \bibinfo {year} {1999})\BibitemShut {NoStop}%
\bibitem [{\citenamefont {Landau}\ and\ \citenamefont
  {Binder}(2014)}]{landau2014guide}%
  \BibitemOpen
  \bibfield  {author} {\bibinfo {author} {\bibfnamefont {D.~P.}\ \bibnamefont
  {Landau}}\ and\ \bibinfo {author} {\bibfnamefont {K.}~\bibnamefont
  {Binder}},\ }\href {https://doi.org/10.1017/CBO9781139696463} {\emph
  {\bibinfo {title} {{A Guide to Monte Carlo Simulations in Statistical
  Physics}}}},\ \bibinfo {edition} {4th}\ ed.\ (\bibinfo  {publisher}
  {Cambridge University Press},\ \bibinfo {year} {2014})\BibitemShut {NoStop}%
\bibitem [{\citenamefont {Metropolis}\ \emph {et~al.}(1953)\citenamefont
  {Metropolis}, \citenamefont {Rosenbluth}, \citenamefont {Rosenbluth},
  \citenamefont {Teller},\ and\ \citenamefont
  {Teller}}]{metropolis1953equation}%
  \BibitemOpen
  \bibfield  {author} {\bibinfo {author} {\bibfnamefont {N.}~\bibnamefont
  {Metropolis}}, \bibinfo {author} {\bibfnamefont {A.~W.}\ \bibnamefont
  {Rosenbluth}}, \bibinfo {author} {\bibfnamefont {M.~N.}\ \bibnamefont
  {Rosenbluth}}, \bibinfo {author} {\bibfnamefont {A.~H.}\ \bibnamefont
  {Teller}},\ and\ \bibinfo {author} {\bibfnamefont {E.}~\bibnamefont
  {Teller}},\ }\bibfield  {title} {\bibinfo {title} {{Equation of state
  calculations by fast computing machines}},\ }\href@noop {} {\bibfield
  {journal} {\bibinfo  {journal} {J. Chem. Phys.}\ }\textbf {\bibinfo {volume}
  {21}},\ \bibinfo {pages} {1087} (\bibinfo {year} {1953})}\BibitemShut
  {NoStop}%
\bibitem [{\citenamefont {Swendsen}\ and\ \citenamefont
  {Wang}(1987)}]{swendsen1987nonuniversal}%
  \BibitemOpen
  \bibfield  {author} {\bibinfo {author} {\bibfnamefont {R.~H.}\ \bibnamefont
  {Swendsen}}\ and\ \bibinfo {author} {\bibfnamefont {J.-S.}\ \bibnamefont
  {Wang}},\ }\bibfield  {title} {\bibinfo {title} {{Nonuniversal critical
  dynamics in {Monte} {Carlo} simulations}},\ }\href
  {https://doi.org/10.1103/PhysRevLett.58.86} {\bibfield  {journal} {\bibinfo
  {journal} {Phys. Rev. Lett.}\ }\textbf {\bibinfo {volume} {58}},\ \bibinfo
  {pages} {86} (\bibinfo {year} {1987})}\BibitemShut {NoStop}%
\bibitem [{\citenamefont {Wolff}(1989)}]{wolff1989collective}%
  \BibitemOpen
  \bibfield  {author} {\bibinfo {author} {\bibfnamefont {U.}~\bibnamefont
  {Wolff}},\ }\bibfield  {title} {\bibinfo {title} {Collective monte carlo
  updating for spin systems},\ }\href
  {https://doi.org/10.1103/PhysRevLett.62.361} {\bibfield  {journal} {\bibinfo
  {journal} {Physical Review Letters}\ }\textbf {\bibinfo {volume} {62}},\
  \bibinfo {pages} {361} (\bibinfo {year} {1989})}\BibitemShut {NoStop}%
\bibitem [{\citenamefont {Fortuin}\ and\ \citenamefont
  {Kasteleyn}(1972)}]{fortuin1972random-cluster}%
  \BibitemOpen
  \bibfield  {author} {\bibinfo {author} {\bibfnamefont {C.~M.}\ \bibnamefont
  {Fortuin}}\ and\ \bibinfo {author} {\bibfnamefont {P.~W.}\ \bibnamefont
  {Kasteleyn}},\ }\bibfield  {title} {\bibinfo {title} {{On the random-cluster
  model: I. Introduction and relation to other models}},\ }\href
  {https://doi.org/10.1016/0031-8914(72)90045-6} {\bibfield  {journal}
  {\bibinfo  {journal} {Physica}\ }\textbf {\bibinfo {volume} {57}},\ \bibinfo
  {pages} {536} (\bibinfo {year} {1972})}\BibitemShut {NoStop}%
\bibitem [{\citenamefont {Zhu}\ \emph {et~al.}(2015)\citenamefont {Zhu},
  \citenamefont {Ochoa},\ and\ \citenamefont {Katzgraber}}]{zhu2015efficient}%
  \BibitemOpen
  \bibfield  {author} {\bibinfo {author} {\bibfnamefont {Z.}~\bibnamefont
  {Zhu}}, \bibinfo {author} {\bibfnamefont {A.~J.}\ \bibnamefont {Ochoa}},\
  and\ \bibinfo {author} {\bibfnamefont {H.~G.}\ \bibnamefont {Katzgraber}},\
  }\bibfield  {title} {\bibinfo {title} {{Efficient {Cluster} {Algorithm} for
  {Spin} {Glasses} in {Any} {Space} {Dimension}}},\ }\href
  {https://doi.org/10.1103/PhysRevLett.115.077201} {\bibfield  {journal}
  {\bibinfo  {journal} {Physical Review Letters}\ }\textbf {\bibinfo {volume}
  {115}},\ \bibinfo {pages} {077201} (\bibinfo {year} {2015})}\BibitemShut
  {NoStop}%
\bibitem [{\citenamefont {Houdayer}(2001)}]{houdayer2001cluster}%
  \BibitemOpen
  \bibfield  {author} {\bibinfo {author} {\bibfnamefont {J.}~\bibnamefont
  {Houdayer}},\ }\bibfield  {title} {\bibinfo {title} {A cluster monte carlo
  algorithm for 2-dimensional spin glasses},\ }\href
  {https://doi.org/10.1007/PL00011151} {\bibfield  {journal} {\bibinfo
  {journal} {The European Physical Journal B - Condensed Matter and Complex
  Systems}\ }\textbf {\bibinfo {volume} {22}},\ \bibinfo {pages} {479}
  (\bibinfo {year} {2001})}\BibitemShut {NoStop}%
\bibitem [{\citenamefont {Schollwöck}(2011)}]{schollwöck2011density-matrix}%
  \BibitemOpen
  \bibfield  {author} {\bibinfo {author} {\bibfnamefont {U.}~\bibnamefont
  {Schollwöck}},\ }\bibfield  {title} {\bibinfo {title} {{The density-matrix
  renormalization group in the age of matrix product states}},\ }\href
  {https://doi.org/10.1016/j.aop.2010.09.012} {\bibfield  {journal} {\bibinfo
  {journal} {Ann. Phys.}\ }\textbf {\bibinfo {volume} {326}},\ \bibinfo {pages}
  {96} (\bibinfo {year} {2011})}\BibitemShut {NoStop}%
\bibitem [{\citenamefont {Orús}(2014)}]{orús2014practical}%
  \BibitemOpen
  \bibfield  {author} {\bibinfo {author} {\bibfnamefont {R.}~\bibnamefont
  {Orús}},\ }\bibfield  {title} {\bibinfo {title} {{A practical introduction
  to tensor networks: {Matrix} product states and projected entangled pair
  states}},\ }\href {https://doi.org/10.1016/j.aop.2014.06.013} {\bibfield
  {journal} {\bibinfo  {journal} {Ann. Phys.}\ }\textbf {\bibinfo {volume}
  {349}},\ \bibinfo {pages} {117} (\bibinfo {year} {2014})}\BibitemShut
  {NoStop}%
\bibitem [{\citenamefont {Ran}\ \emph {et~al.}(2020)\citenamefont {Ran},
  \citenamefont {Tirrito}, \citenamefont {Peng}, \citenamefont {Chen},
  \citenamefont {Tagliacozzo}, \citenamefont {Su},\ and\ \citenamefont
  {Lewenstein}}]{ran2020tensor}%
  \BibitemOpen
  \bibfield  {author} {\bibinfo {author} {\bibfnamefont {S.-J.}\ \bibnamefont
  {Ran}}, \bibinfo {author} {\bibfnamefont {E.}~\bibnamefont {Tirrito}},
  \bibinfo {author} {\bibfnamefont {C.}~\bibnamefont {Peng}}, \bibinfo {author}
  {\bibfnamefont {X.}~\bibnamefont {Chen}}, \bibinfo {author} {\bibfnamefont
  {L.}~\bibnamefont {Tagliacozzo}}, \bibinfo {author} {\bibfnamefont
  {G.}~\bibnamefont {Su}},\ and\ \bibinfo {author} {\bibfnamefont
  {M.}~\bibnamefont {Lewenstein}},\ }\href
  {https://doi.org/10.1007/978-3-030-34489-4} {\emph {\bibinfo {title} {{Tensor
  Network Contractions: Methods and Applications to Quantum Many-Body
  Systems}}}}\ (\bibinfo  {publisher} {Springer International Publishing},\
  \bibinfo {year} {2020})\BibitemShut {NoStop}%
\bibitem [{\citenamefont {Cirac}\ \emph {et~al.}(2021)\citenamefont {Cirac},
  \citenamefont {P\'erez-Garc\'{\i}a}, \citenamefont {Schuch},\ and\
  \citenamefont {Verstraete}}]{cirac2021matrix}%
  \BibitemOpen
  \bibfield  {author} {\bibinfo {author} {\bibfnamefont {J.~I.}\ \bibnamefont
  {Cirac}}, \bibinfo {author} {\bibfnamefont {D.}~\bibnamefont
  {P\'erez-Garc\'{\i}a}}, \bibinfo {author} {\bibfnamefont {N.}~\bibnamefont
  {Schuch}},\ and\ \bibinfo {author} {\bibfnamefont {F.}~\bibnamefont
  {Verstraete}},\ }\bibfield  {title} {\bibinfo {title} {{Matrix product states
  and projected entangled pair states: {Concepts}, symmetries, theorems}},\
  }\href {https://doi.org/10.1103/RevModPhys.93.045003} {\bibfield  {journal}
  {\bibinfo  {journal} {Rev. Mod. Phys.}\ }\textbf {\bibinfo {volume} {93}},\
  \bibinfo {pages} {045003} (\bibinfo {year} {2021})}\BibitemShut {NoStop}%
\bibitem [{\citenamefont {Pan}\ and\ \citenamefont
  {Zhang}(2022)}]{pan2022simulation}%
  \BibitemOpen
  \bibfield  {author} {\bibinfo {author} {\bibfnamefont {F.}~\bibnamefont
  {Pan}}\ and\ \bibinfo {author} {\bibfnamefont {P.}~\bibnamefont {Zhang}},\
  }\bibfield  {title} {\bibinfo {title} {Simulation of quantum circuits using
  the big-batch tensor network method},\ }\href
  {https://journals.aps.org/prl/abstract/10.1103/PhysRevLett.128.030501}
  {\bibfield  {journal} {\bibinfo  {journal} {Physical Review Letters}\
  }\textbf {\bibinfo {volume} {128}},\ \bibinfo {pages} {030501} (\bibinfo
  {year} {2022})}\BibitemShut {NoStop}%
\bibitem [{\citenamefont {Chen}\ \emph
  {et~al.}(2025{\natexlab{a}})\citenamefont {Chen}, \citenamefont {Guo},
  \citenamefont {Zhang}, \citenamefont {Zhang},\ and\ \citenamefont
  {Deng}}]{chen2025tensor}%
  \BibitemOpen
  \bibfield  {author} {\bibinfo {author} {\bibfnamefont {T.}~\bibnamefont
  {Chen}}, \bibinfo {author} {\bibfnamefont {E.}~\bibnamefont {Guo}}, \bibinfo
  {author} {\bibfnamefont {W.}~\bibnamefont {Zhang}}, \bibinfo {author}
  {\bibfnamefont {P.}~\bibnamefont {Zhang}},\ and\ \bibinfo {author}
  {\bibfnamefont {Y.}~\bibnamefont {Deng}},\ }\bibfield  {title} {\bibinfo
  {title} {{Tensor network {Monte} {Carlo} simulations for the two-dimensional
  random-bond {Ising} model}},\ }\href
  {https://doi.org/10.1103/PhysRevB.111.094201} {\bibfield  {journal} {\bibinfo
   {journal} {Physical Review B}\ }\textbf {\bibinfo {volume} {111}},\ \bibinfo
  {pages} {094201} (\bibinfo {year} {2025}{\natexlab{a}})}\BibitemShut
  {NoStop}%
\bibitem [{\citenamefont {Pan}\ \emph {et~al.}(2020)\citenamefont {Pan},
  \citenamefont {Zhou}, \citenamefont {Li},\ and\ \citenamefont
  {Zhang}}]{pan2020contracting}%
  \BibitemOpen
  \bibfield  {author} {\bibinfo {author} {\bibfnamefont {F.}~\bibnamefont
  {Pan}}, \bibinfo {author} {\bibfnamefont {P.}~\bibnamefont {Zhou}}, \bibinfo
  {author} {\bibfnamefont {S.}~\bibnamefont {Li}},\ and\ \bibinfo {author}
  {\bibfnamefont {P.}~\bibnamefont {Zhang}},\ }\bibfield  {title} {\bibinfo
  {title} {Contracting arbitrary tensor networks: general approximate algorithm
  and applications in graphical models and quantum circuit simulations},\
  }\href {https://journals.aps.org/prl/abstract/10.1103/PhysRevLett.125.060503}
  {\bibfield  {journal} {\bibinfo  {journal} {Physical Review Letters}\
  }\textbf {\bibinfo {volume} {125}},\ \bibinfo {pages} {060503} (\bibinfo
  {year} {2020})}\BibitemShut {NoStop}%
\bibitem [{\citenamefont {Liu}\ \emph {et~al.}(2021{\natexlab{a}})\citenamefont
  {Liu}, \citenamefont {Wang},\ and\ \citenamefont {Zhang}}]{liu2021tropical}%
  \BibitemOpen
  \bibfield  {author} {\bibinfo {author} {\bibfnamefont {J.-G.}\ \bibnamefont
  {Liu}}, \bibinfo {author} {\bibfnamefont {L.}~\bibnamefont {Wang}},\ and\
  \bibinfo {author} {\bibfnamefont {P.}~\bibnamefont {Zhang}},\ }\bibfield
  {title} {\bibinfo {title} {{{Tropical} {Tensor} {Network} for {Ground}
  {States} of {Spin} {Glasses}}},\ }\href
  {https://doi.org/10.1103/PhysRevLett.126.090506} {\bibfield  {journal}
  {\bibinfo  {journal} {Phys. Rev. Lett.}\ }\textbf {\bibinfo {volume} {126}},\
  \bibinfo {pages} {090506} (\bibinfo {year} {2021}{\natexlab{a}})}\BibitemShut
  {NoStop}%
\bibitem [{\citenamefont {Okunishi}\ \emph {et~al.}(2022)\citenamefont
  {Okunishi}, \citenamefont {Nishino},\ and\ \citenamefont
  {Ueda}}]{okunishi2022developments}%
  \BibitemOpen
  \bibfield  {author} {\bibinfo {author} {\bibfnamefont {K.}~\bibnamefont
  {Okunishi}}, \bibinfo {author} {\bibfnamefont {T.}~\bibnamefont {Nishino}},\
  and\ \bibinfo {author} {\bibfnamefont {H.}~\bibnamefont {Ueda}},\ }\bibfield
  {title} {\bibinfo {title} {{{Developments} in the {Tensor} {Network} — from
  {Statistical} {Mechanics} to {Quantum} {Entanglement}}},\ }\href
  {https://doi.org/10.7566/JPSJ.91.062001} {\bibfield  {journal} {\bibinfo
  {journal} {J. Phys. Soc. Jpn.}\ }\textbf {\bibinfo {volume} {91}},\ \bibinfo
  {pages} {062001} (\bibinfo {year} {2022})}\BibitemShut {NoStop}%
\bibitem [{\citenamefont {White}(1992)}]{white1992density}%
  \BibitemOpen
  \bibfield  {author} {\bibinfo {author} {\bibfnamefont {S.~R.}\ \bibnamefont
  {White}},\ }\bibfield  {title} {\bibinfo {title} {{Density matrix formulation
  for quantum renormalization groups}},\ }\href
  {https://doi.org/10.1103/PhysRevLett.69.2863} {\bibfield  {journal} {\bibinfo
   {journal} {Phys. Rev. Lett.}\ }\textbf {\bibinfo {volume} {69}},\ \bibinfo
  {pages} {2863} (\bibinfo {year} {1992})}\BibitemShut {NoStop}%
\bibitem [{\citenamefont {Nishino}\ and\ \citenamefont
  {Okunishi}(1998)}]{nishino1998density}%
  \BibitemOpen
  \bibfield  {author} {\bibinfo {author} {\bibfnamefont {T.}~\bibnamefont
  {Nishino}}\ and\ \bibinfo {author} {\bibfnamefont {K.}~\bibnamefont
  {Okunishi}},\ }\bibfield  {title} {\bibinfo {title} {{A {Density} {Matrix}
  {Algorithm} for {3D} {Classical} {Models}}},\ }\href
  {https://doi.org/10.1143/JPSJ.67.3066} {\bibfield  {journal} {\bibinfo
  {journal} {J. Phys. Soc. Jpn.}\ }\textbf {\bibinfo {volume} {67}},\ \bibinfo
  {pages} {3066} (\bibinfo {year} {1998})}\BibitemShut {NoStop}%
\bibitem [{\citenamefont {Chen}\ \emph {et~al.}(2011)\citenamefont {Chen},
  \citenamefont {Qin}, \citenamefont {Chen}, \citenamefont {Wei}, \citenamefont
  {Zhao}, \citenamefont {Normand},\ and\ \citenamefont
  {Xiang}}]{chen2011partial}%
  \BibitemOpen
  \bibfield  {author} {\bibinfo {author} {\bibfnamefont {Q.~N.}\ \bibnamefont
  {Chen}}, \bibinfo {author} {\bibfnamefont {M.~P.}\ \bibnamefont {Qin}},
  \bibinfo {author} {\bibfnamefont {J.}~\bibnamefont {Chen}}, \bibinfo {author}
  {\bibfnamefont {Z.~C.}\ \bibnamefont {Wei}}, \bibinfo {author} {\bibfnamefont
  {H.~H.}\ \bibnamefont {Zhao}}, \bibinfo {author} {\bibfnamefont
  {B.}~\bibnamefont {Normand}},\ and\ \bibinfo {author} {\bibfnamefont
  {T.}~\bibnamefont {Xiang}},\ }\bibfield  {title} {\bibinfo {title}
  {{{Partial} {Order} and {Finite-Temperature} {Phase} {Transitions} in {Potts}
  {Models} on {Irregular} {Lattices}}},\ }\href
  {https://doi.org/10.1103/PhysRevLett.107.165701} {\bibfield  {journal}
  {\bibinfo  {journal} {Phys. Rev. Lett.}\ }\textbf {\bibinfo {volume} {107}},\
  \bibinfo {pages} {165701} (\bibinfo {year} {2011})}\BibitemShut {NoStop}%
\bibitem [{\citenamefont {Xie}\ \emph {et~al.}(2009)\citenamefont {Xie},
  \citenamefont {Jiang}, \citenamefont {Chen}, \citenamefont {Weng},\ and\
  \citenamefont {Xiang}}]{xie2009second}%
  \BibitemOpen
  \bibfield  {author} {\bibinfo {author} {\bibfnamefont {Z.~Y.}\ \bibnamefont
  {Xie}}, \bibinfo {author} {\bibfnamefont {H.~C.}\ \bibnamefont {Jiang}},
  \bibinfo {author} {\bibfnamefont {Q.~N.}\ \bibnamefont {Chen}}, \bibinfo
  {author} {\bibfnamefont {Z.~Y.}\ \bibnamefont {Weng}},\ and\ \bibinfo
  {author} {\bibfnamefont {T.}~\bibnamefont {Xiang}},\ }\bibfield  {title}
  {\bibinfo {title} {{Second Renormalization of Tensor-Network States}},\
  }\href {https://doi.org/10.1103/PhysRevLett.103.160601} {\bibfield  {journal}
  {\bibinfo  {journal} {Phys. Rev. Lett.}\ }\textbf {\bibinfo {volume} {103}},\
  \bibinfo {pages} {160601} (\bibinfo {year} {2009})}\BibitemShut {NoStop}%
\bibitem [{\citenamefont {Liu}\ \emph {et~al.}(2022)\citenamefont {Liu},
  \citenamefont {Fu}, \citenamefont {Yu}, \citenamefont {Yu},\ and\
  \citenamefont {Xie}}]{fliu2022variational}%
  \BibitemOpen
  \bibfield  {author} {\bibinfo {author} {\bibfnamefont {X.~F.}\ \bibnamefont
  {Liu}}, \bibinfo {author} {\bibfnamefont {Y.~F.}\ \bibnamefont {Fu}},
  \bibinfo {author} {\bibfnamefont {W.~Q.}\ \bibnamefont {Yu}}, \bibinfo
  {author} {\bibfnamefont {J.~F.}\ \bibnamefont {Yu}},\ and\ \bibinfo {author}
  {\bibfnamefont {Z.~Y.}\ \bibnamefont {Xie}},\ }\bibfield  {title} {\bibinfo
  {title} {{{Variational} {Corner} {Transfer} {Matrix} {Renormalization}
  {Group} {Method} for {Classical} {Statistical} {Models}}},\ }\href
  {https://doi.org/10.1088/0256-307X/39/6/067502} {\bibfield  {journal}
  {\bibinfo  {journal} {Chin. Phys. Lett.}\ }\textbf {\bibinfo {volume} {39}},\
  \bibinfo {pages} {067502} (\bibinfo {year} {2022})}\BibitemShut {NoStop}%
\bibitem [{\citenamefont {Xiang}(2023)}]{xiang2023density}%
  \BibitemOpen
  \bibfield  {author} {\bibinfo {author} {\bibfnamefont {T.}~\bibnamefont
  {Xiang}},\ }\href@noop {} {\emph {\bibinfo {title} {{Density Matrix and
  Tensor Network Renormalization}}}}\ (\bibinfo  {publisher} {Cambridge
  University Press},\ \bibinfo {year} {2023})\BibitemShut {NoStop}%
\bibitem [{\citenamefont {Baity-Jesi}\ \emph {et~al.}(2013)\citenamefont
  {Baity-Jesi}, \citenamefont {Baños}, \citenamefont {Cruz}, \citenamefont
  {Fernandez}, \citenamefont {Gil-Narvion}, \citenamefont {Gordillo-Guerrero},
  \citenamefont {Iñiguez}, \citenamefont {Maiorano}, \citenamefont
  {Mantovani}, \citenamefont {Marinari}, \citenamefont {Martin-Mayor},
  \citenamefont {Monforte-Garcia}, \citenamefont {Sudupe}, \citenamefont
  {Navarro}, \citenamefont {Parisi}, \citenamefont {Perez-Gaviro},
  \citenamefont {Pivanti}, \citenamefont {Ricci-Tersenghi}, \citenamefont
  {Ruiz-Lorenzo}, \citenamefont {Schifano}, \citenamefont {Seoane},
  \citenamefont {Tarancon}, \citenamefont {Tripiccione}, \citenamefont
  {Yllanes},\ and\ \citenamefont {{Janus
  Collaboration}}}]{baity-jesi2013critical}%
  \BibitemOpen
  \bibfield  {author} {\bibinfo {author} {\bibfnamefont {M.}~\bibnamefont
  {Baity-Jesi}}, \bibinfo {author} {\bibfnamefont {R.~A.}\ \bibnamefont
  {Baños}}, \bibinfo {author} {\bibfnamefont {A.}~\bibnamefont {Cruz}},
  \bibinfo {author} {\bibfnamefont {L.~A.}\ \bibnamefont {Fernandez}}, \bibinfo
  {author} {\bibfnamefont {J.~M.}\ \bibnamefont {Gil-Narvion}}, \bibinfo
  {author} {\bibfnamefont {A.}~\bibnamefont {Gordillo-Guerrero}}, \bibinfo
  {author} {\bibfnamefont {D.}~\bibnamefont {Iñiguez}}, \bibinfo {author}
  {\bibfnamefont {A.}~\bibnamefont {Maiorano}}, \bibinfo {author}
  {\bibfnamefont {F.}~\bibnamefont {Mantovani}}, \bibinfo {author}
  {\bibfnamefont {E.}~\bibnamefont {Marinari}}, \bibinfo {author}
  {\bibfnamefont {V.}~\bibnamefont {Martin-Mayor}}, \bibinfo {author}
  {\bibfnamefont {J.}~\bibnamefont {Monforte-Garcia}}, \bibinfo {author}
  {\bibfnamefont {A.~M.}\ \bibnamefont {Sudupe}}, \bibinfo {author}
  {\bibfnamefont {D.}~\bibnamefont {Navarro}}, \bibinfo {author} {\bibfnamefont
  {G.}~\bibnamefont {Parisi}}, \bibinfo {author} {\bibfnamefont
  {S.}~\bibnamefont {Perez-Gaviro}}, \bibinfo {author} {\bibfnamefont
  {M.}~\bibnamefont {Pivanti}}, \bibinfo {author} {\bibfnamefont
  {F.}~\bibnamefont {Ricci-Tersenghi}}, \bibinfo {author} {\bibfnamefont
  {J.~J.}\ \bibnamefont {Ruiz-Lorenzo}}, \bibinfo {author} {\bibfnamefont
  {S.~F.}\ \bibnamefont {Schifano}}, \bibinfo {author} {\bibfnamefont
  {B.}~\bibnamefont {Seoane}}, \bibinfo {author} {\bibfnamefont
  {A.}~\bibnamefont {Tarancon}}, \bibinfo {author} {\bibfnamefont
  {R.}~\bibnamefont {Tripiccione}}, \bibinfo {author} {\bibfnamefont
  {D.}~\bibnamefont {Yllanes}},\ and\ \bibinfo {author} {\bibnamefont {{Janus
  Collaboration}}},\ }\bibfield  {title} {\bibinfo {title} {{Critical
  parameters of the three-dimensional {Ising} spin glass}},\ }\href
  {https://doi.org/10.1103/PhysRevB.88.224416} {\bibfield  {journal} {\bibinfo
  {journal} {Physical Review B}\ }\textbf {\bibinfo {volume} {88}},\ \bibinfo
  {pages} {224416} (\bibinfo {year} {2013})}\BibitemShut {NoStop}%
\bibitem [{\citenamefont {Janke}\ and\ \citenamefont
  {Villanova}(1997)}]{janke1997three-dimensional}%
  \BibitemOpen
  \bibfield  {author} {\bibinfo {author} {\bibfnamefont {W.}~\bibnamefont
  {Janke}}\ and\ \bibinfo {author} {\bibfnamefont {R.}~\bibnamefont
  {Villanova}},\ }\bibfield  {title} {\bibinfo {title} {{Three-dimensional
  3-state {Potts} model revisited with new techniques}},\ }\href
  {https://doi.org/10.1016/S0550-3213(96)00710-9} {\bibfield  {journal}
  {\bibinfo  {journal} {Nuclear Physics B}\ }\textbf {\bibinfo {volume}
  {489}},\ \bibinfo {pages} {679} (\bibinfo {year} {1997})}\BibitemShut
  {NoStop}%
\bibitem [{\citenamefont {Liu}\ \emph {et~al.}(2021{\natexlab{b}})\citenamefont
  {Liu}, \citenamefont {Huang}, \citenamefont {Gong},\ and\ \citenamefont
  {Gu}}]{liu2021accurate}%
  \BibitemOpen
  \bibfield  {author} {\bibinfo {author} {\bibfnamefont {W.-Y.}\ \bibnamefont
  {Liu}}, \bibinfo {author} {\bibfnamefont {Y.-Z.}\ \bibnamefont {Huang}},
  \bibinfo {author} {\bibfnamefont {S.-S.}\ \bibnamefont {Gong}},\ and\
  \bibinfo {author} {\bibfnamefont {Z.-C.}\ \bibnamefont {Gu}},\ }\bibfield
  {title} {\bibinfo {title} {Accurate simulation for finite projected entangled
  pair states in two dimensions},\ }\href
  {https://doi.org/10.1103/PhysRevB.103.235155} {\bibfield  {journal} {\bibinfo
   {journal} {Physical Review B}\ }\textbf {\bibinfo {volume} {103}},\ \bibinfo
  {pages} {235155} (\bibinfo {year} {2021}{\natexlab{b}})}\BibitemShut
  {NoStop}%
\bibitem [{\citenamefont {Liu}\ \emph {et~al.}(2025)\citenamefont {Liu},
  \citenamefont {Zhai}, \citenamefont {Peng}, \citenamefont {Gu},\ and\
  \citenamefont {Chan}}]{liu2025accurate}%
  \BibitemOpen
  \bibfield  {author} {\bibinfo {author} {\bibfnamefont {W.-Y.}\ \bibnamefont
  {Liu}}, \bibinfo {author} {\bibfnamefont {H.}~\bibnamefont {Zhai}}, \bibinfo
  {author} {\bibfnamefont {R.}~\bibnamefont {Peng}}, \bibinfo {author}
  {\bibfnamefont {Z.-C.}\ \bibnamefont {Gu}},\ and\ \bibinfo {author}
  {\bibfnamefont {G.~K.-L.}\ \bibnamefont {Chan}},\ }\bibfield  {title}
  {\bibinfo {title} {Accurate simulation of the hubbard model with finite
  fermionic projected entangled pair states},\ }\href
  {https://doi.org/10.1103/r4q9-4yvj} {\bibfield  {journal} {\bibinfo
  {journal} {Physical Review Letters}\ }\textbf {\bibinfo {volume} {134}},\
  \bibinfo {pages} {256502} (\bibinfo {year} {2025})}\BibitemShut {NoStop}%
\bibitem [{\citenamefont {Layden}\ \emph {et~al.}(2023)\citenamefont {Layden},
  \citenamefont {Mazzola}, \citenamefont {Mishmash}, \citenamefont {Motta},
  \citenamefont {Wocjan}, \citenamefont {Kim},\ and\ \citenamefont
  {Sheldon}}]{layden2023quantumenhanced}%
  \BibitemOpen
  \bibfield  {author} {\bibinfo {author} {\bibfnamefont {D.}~\bibnamefont
  {Layden}}, \bibinfo {author} {\bibfnamefont {G.}~\bibnamefont {Mazzola}},
  \bibinfo {author} {\bibfnamefont {R.~V.}\ \bibnamefont {Mishmash}}, \bibinfo
  {author} {\bibfnamefont {M.}~\bibnamefont {Motta}}, \bibinfo {author}
  {\bibfnamefont {P.}~\bibnamefont {Wocjan}}, \bibinfo {author} {\bibfnamefont
  {J.-S.}\ \bibnamefont {Kim}},\ and\ \bibinfo {author} {\bibfnamefont
  {S.}~\bibnamefont {Sheldon}},\ }\bibfield  {title} {\bibinfo {title}
  {Quantum-enhanced markov chain monte carlo},\ }\href
  {https://doi.org/10.1038/s41586-023-06095-4} {\bibfield  {journal} {\bibinfo
  {journal} {Nature}\ }\textbf {\bibinfo {volume} {619}},\ \bibinfo {pages}
  {282} (\bibinfo {year} {2023})}\BibitemShut {NoStop}%
\bibitem [{tnm()}]{tnmcmc-repo}%
  \BibitemOpen
  \href@noop {} {}\bibinfo {howpublished}
  {\url{https://github.com/Fermichen99/TNMCMC}}\BibitemShut {NoStop}%
\bibitem [{\citenamefont {Weigel}(2012)}]{weigel2012performance}%
  \BibitemOpen
  \bibfield  {author} {\bibinfo {author} {\bibfnamefont {M.}~\bibnamefont
  {Weigel}},\ }\bibfield  {title} {\bibinfo {title} {Performance potential for
  simulating spin models on gpu},\ }\href
  {https://doi.org/10.1016/j.jcp.2011.12.008} {\bibfield  {journal} {\bibinfo
  {journal} {Journal of Computational Physics}\ }\textbf {\bibinfo {volume}
  {231}},\ \bibinfo {pages} {3064} (\bibinfo {year} {2012})}\BibitemShut
  {NoStop}%
\bibitem [{\citenamefont {Bisson}\ \emph
  {et~al.}(2025{\natexlab{a}})\citenamefont {Bisson}, \citenamefont
  {Bernaschi}, \citenamefont {Fatica}, \citenamefont {Fytas}, \citenamefont
  {{Gonz{\'a}lez-Adalid Pemart{\'i}n}}, \citenamefont {{Mart{\'i}n-Mayor}},\
  and\ \citenamefont {Vasilopoulos}}]{bisson2025massivescale}%
  \BibitemOpen
  \bibfield  {author} {\bibinfo {author} {\bibfnamefont {M.}~\bibnamefont
  {Bisson}}, \bibinfo {author} {\bibfnamefont {M.}~\bibnamefont {Bernaschi}},
  \bibinfo {author} {\bibfnamefont {M.}~\bibnamefont {Fatica}}, \bibinfo
  {author} {\bibfnamefont {N.~G.}\ \bibnamefont {Fytas}}, \bibinfo {author}
  {\bibfnamefont {I.}~\bibnamefont {{Gonz{\'a}lez-Adalid Pemart{\'i}n}}},
  \bibinfo {author} {\bibfnamefont {V.}~\bibnamefont {{Mart{\'i}n-Mayor}}},\
  and\ \bibinfo {author} {\bibfnamefont {A.}~\bibnamefont {Vasilopoulos}},\
  }\bibfield  {title} {\bibinfo {title} {Massive-scale simulations of 2d ising
  and blume-capel models on rack-scale multi-gpu systems},\ }\href
  {https://doi.org/10.1016/j.cpc.2025.109690} {\bibfield  {journal} {\bibinfo
  {journal} {Computer Physics Communications}\ }\textbf {\bibinfo {volume}
  {315}},\ \bibinfo {pages} {109690} (\bibinfo {year}
  {2025}{\natexlab{a}})}\BibitemShut {NoStop}%
\bibitem [{\citenamefont {Bisson}\ \emph
  {et~al.}(2025{\natexlab{b}})\citenamefont {Bisson}, \citenamefont
  {Vasilopoulos}, \citenamefont {Bernaschi}, \citenamefont {Fatica},
  \citenamefont {Fytas}, \citenamefont {Pemart{\'i}n},\ and\ \citenamefont
  {{Mart{\'i}n-Mayor}}}]{bisson2025universal}%
  \BibitemOpen
  \bibfield  {author} {\bibinfo {author} {\bibfnamefont {M.}~\bibnamefont
  {Bisson}}, \bibinfo {author} {\bibfnamefont {A.}~\bibnamefont
  {Vasilopoulos}}, \bibinfo {author} {\bibfnamefont {M.}~\bibnamefont
  {Bernaschi}}, \bibinfo {author} {\bibfnamefont {M.}~\bibnamefont {Fatica}},
  \bibinfo {author} {\bibfnamefont {N.~G.}\ \bibnamefont {Fytas}}, \bibinfo
  {author} {\bibfnamefont {I.~G.-A.}\ \bibnamefont {Pemart{\'i}n}},\ and\
  \bibinfo {author} {\bibfnamefont {V.}~\bibnamefont {{Mart{\'i}n-Mayor}}},\
  }\bibfield  {title} {\bibinfo {title} {Universal exotic dynamics in critical
  mesoscopic systems: Simulating the square root of avogadro's number of
  spins},\ }\href {https://doi.org/10.1103/ngkf-7816} {\bibfield  {journal}
  {\bibinfo  {journal} {Physical Review Research}\ }\textbf {\bibinfo {volume}
  {7}},\ \bibinfo {pages} {033218} (\bibinfo {year}
  {2025}{\natexlab{b}})}\BibitemShut {NoStop}%
\bibitem [{\citenamefont {Chen}\ \emph
  {et~al.}(2025{\natexlab{b}})\citenamefont {Chen}, \citenamefont {Zhang},
  \citenamefont {Liu}, \citenamefont {Deng},\ and\ \citenamefont
  {Zhang}}]{chen2025batchtnmc}%
  \BibitemOpen
  \bibfield  {author} {\bibinfo {author} {\bibfnamefont {T.}~\bibnamefont
  {Chen}}, \bibinfo {author} {\bibfnamefont {J.}~\bibnamefont {Zhang}},
  \bibinfo {author} {\bibfnamefont {J.}~\bibnamefont {Liu}}, \bibinfo {author}
  {\bibfnamefont {Y.}~\bibnamefont {Deng}},\ and\ \bibinfo {author}
  {\bibfnamefont {P.}~\bibnamefont {Zhang}},\ }\href
  {https://arxiv.org/abs/2509.19006} {\bibinfo {title} {Batchtnmc: Efficient
  sampling of two-dimensional spin glasses using tensor network monte carlo}}
  (\bibinfo {year} {2025}{\natexlab{b}}),\ \Eprint
  {https://arxiv.org/abs/2509.19006} {arXiv:2509.19006 [cond-mat.stat-mech]}
  \BibitemShut {NoStop}%
\end{thebibliography}%


\begin{thebibliography}{6}%
\makeatletter
\providecommand \@ifxundefined [1]{%
 \@ifx{#1\undefined}
}%
\providecommand \@ifnum [1]{%
 \ifnum #1\expandafter \@firstoftwo
 \else \expandafter \@secondoftwo
 \fi
}%
\providecommand \@ifx [1]{%
 \ifx #1\expandafter \@firstoftwo
 \else \expandafter \@secondoftwo
 \fi
}%
\providecommand \natexlab [1]{#1}%
\providecommand \enquote  [1]{``#1''}%
\providecommand \bibnamefont  [1]{#1}%
\providecommand \bibfnamefont [1]{#1}%
\providecommand \citenamefont [1]{#1}%
\providecommand \href@noop [0]{\@secondoftwo}%
\providecommand \href [0]{\begingroup \@sanitize@url \@href}%
\providecommand \@href[1]{\@@startlink{#1}\@@href}%
\providecommand \@@href[1]{\endgroup#1\@@endlink}%
\providecommand \@sanitize@url [0]{\catcode `\\12\catcode `\$12\catcode
  `\&12\catcode `\#12\catcode `\^12\catcode `\_12\catcode `\%12\relax}%
\providecommand \@@startlink[1]{}%
\providecommand \@@endlink[0]{}%
\providecommand \url  [0]{\begingroup\@sanitize@url \@url }%
\providecommand \@url [1]{\endgroup\@href {#1}{\urlprefix }}%
\providecommand \urlprefix  [0]{URL }%
\providecommand \Eprint [0]{\href }%
\providecommand \doibase [0]{https://doi.org/}%
\providecommand \selectlanguage [0]{\@gobble}%
\providecommand \bibinfo  [0]{\@secondoftwo}%
\providecommand \bibfield  [0]{\@secondoftwo}%
\providecommand \translation [1]{[#1]}%
\providecommand \BibitemOpen [0]{}%
\providecommand \bibitemStop [0]{}%
\providecommand \bibitemNoStop [0]{.\EOS\space}%
\providecommand \EOS [0]{\spacefactor3000\relax}%
\providecommand \BibitemShut  [1]{\csname bibitem#1\endcsname}%
\let\auto@bib@innerbib\@empty
\bibitem [{\citenamefont {Chen}\ \emph {et~al.}(2025)\citenamefont {Chen},
  \citenamefont {Guo}, \citenamefont {Zhang}, \citenamefont {Zhang},\ and\
  \citenamefont {Deng}}]{chen2025tensor}%
  \BibitemOpen
  \bibfield  {author} {\bibinfo {author} {\bibfnamefont {T.}~\bibnamefont
  {Chen}}, \bibinfo {author} {\bibfnamefont {E.}~\bibnamefont {Guo}}, \bibinfo
  {author} {\bibfnamefont {W.}~\bibnamefont {Zhang}}, \bibinfo {author}
  {\bibfnamefont {P.}~\bibnamefont {Zhang}},\ and\ \bibinfo {author}
  {\bibfnamefont {Y.}~\bibnamefont {Deng}},\ }\bibfield  {title} {\bibinfo
  {title} {{Tensor network {Monte} {Carlo} simulations for the two-dimensional
  random-bond {Ising} model}},\ }\href
  {https://doi.org/10.1103/PhysRevB.111.094201} {\bibfield  {journal} {\bibinfo
   {journal} {Physical Review B}\ }\textbf {\bibinfo {volume} {111}},\ \bibinfo
  {pages} {094201} (\bibinfo {year} {2025})}\BibitemShut {NoStop}%
\bibitem [{\citenamefont {Levin}\ and\ \citenamefont
  {Nave}(2007)}]{levin2007tensor}%
  \BibitemOpen
  \bibfield  {author} {\bibinfo {author} {\bibfnamefont {M.}~\bibnamefont
  {Levin}}\ and\ \bibinfo {author} {\bibfnamefont {C.~P.}\ \bibnamefont
  {Nave}},\ }\bibfield  {title} {\bibinfo {title} {{{Tensor} {Renormalization}
  {Group} {Approach} to {Two-Dimensional} {Classical} {Lattice} {Models}}},\
  }\href {https://doi.org/10.1103/PhysRevLett.99.120601} {\bibfield  {journal}
  {\bibinfo  {journal} {Phys. Rev. Lett.}\ }\textbf {\bibinfo {volume} {99}},\
  \bibinfo {pages} {120601} (\bibinfo {year} {2007})}\BibitemShut {NoStop}%
\bibitem [{\citenamefont {Orús}(2014)}]{orús2014practical}%
  \BibitemOpen
  \bibfield  {author} {\bibinfo {author} {\bibfnamefont {R.}~\bibnamefont
  {Orús}},\ }\bibfield  {title} {\bibinfo {title} {{A practical introduction
  to tensor networks: {Matrix} product states and projected entangled pair
  states}},\ }\href {https://doi.org/10.1016/j.aop.2014.06.013} {\bibfield
  {journal} {\bibinfo  {journal} {Ann. Phys.}\ }\textbf {\bibinfo {volume}
  {349}},\ \bibinfo {pages} {117} (\bibinfo {year} {2014})}\BibitemShut
  {NoStop}%
\bibitem [{\citenamefont {Bishop}(2006)}]{bishop2006pattern}%
  \BibitemOpen
  \bibfield  {author} {\bibinfo {author} {\bibfnamefont {C.~M.}\ \bibnamefont
  {Bishop}},\ }\href@noop {} {\emph {\bibinfo {title} {Pattern Recognition and
  Machine Learning}}}\ (\bibinfo  {publisher} {Springer New York, NY},\
  \bibinfo {year} {2006})\BibitemShut {NoStop}%
\bibitem [{\citenamefont {Swendsen}\ and\ \citenamefont
  {Wang}(1987)}]{swendsen1987nonuniversal}%
  \BibitemOpen
  \bibfield  {author} {\bibinfo {author} {\bibfnamefont {R.~H.}\ \bibnamefont
  {Swendsen}}\ and\ \bibinfo {author} {\bibfnamefont {J.-S.}\ \bibnamefont
  {Wang}},\ }\bibfield  {title} {\bibinfo {title} {{Nonuniversal critical
  dynamics in {Monte} {Carlo} simulations}},\ }\href
  {https://doi.org/10.1103/PhysRevLett.58.86} {\bibfield  {journal} {\bibinfo
  {journal} {Phys. Rev. Lett.}\ }\textbf {\bibinfo {volume} {58}},\ \bibinfo
  {pages} {86} (\bibinfo {year} {1987})}\BibitemShut {NoStop}%
\bibitem [{\citenamefont {Wolff}(1989)}]{wolff1989collective}%
  \BibitemOpen
  \bibfield  {author} {\bibinfo {author} {\bibfnamefont {U.}~\bibnamefont
  {Wolff}},\ }\bibfield  {title} {\bibinfo {title} {Collective monte carlo
  updating for spin systems},\ }\href
  {https://doi.org/10.1103/PhysRevLett.62.361} {\bibfield  {journal} {\bibinfo
  {journal} {Physical Review Letters}\ }\textbf {\bibinfo {volume} {62}},\
  \bibinfo {pages} {361} (\bibinfo {year} {1989})}\BibitemShut {NoStop}%
\end{thebibliography}%

\end{document}


\title{\textit{Supplemental Material of} Tensor Network Markov Chain Monte Carlo (TNMCMC): Efficient Sampling of Three-Dimensional Spin Glasses and Beyond}

\maketitle
\tableofcontents

\section{Tensor Network Monte Carlo (TNMC) method for 3D systems}
\label{section:sm-tnmc}

This section provides a detailed derivation of the tensor-network representation of the partition function and outlines the procedures of the Tensor Network Monte Carlo (TNMC) algorithm for 3D systems.
As for 2D systems, see~\cite{chen2025tensor} for more details.

\subsection{Tensor network representation of the 3D Ising partition function}
\label{subsection:sm-tnmc-tensor-network}

Consider the classical 3D Ising Hamiltonian on a cubic lattice: 
\begin{equation}
E(\s) = -\sum_{\langle i,j\rangle} J_{ij} s_i s_j, \quad s_i \in \{+1,-1\},
\end{equation}
the partition function is given by
\begin{equation}
    Z = \sum_{\s} \exp\bigl(-\beta E(\s)\bigr).
\end{equation}
Following established methods~\cite{levin2007tensor,orús2014practical}, one may rewrite the Boltzmann weight for each nearest-neighbor interaction as a matrix element:
\begin{equation}
    W_{ij} =
        \begin{pmatrix}
            e^{\beta J_{ij}} & e^{-\beta J_{ij}}\\
            e^{-\beta J_{ij}} & e^{\beta J_{ij}}
        \end{pmatrix},
\end{equation}
and at each lattice node, we introduce a rank-6 copy tensor with indices corresponding to the six adjacent bonds:
\begin{equation}
    \delta_{iakbmc}=\begin{cases}
1, & \text{if } i=a=k=b=m=c, \\ 0, & \text{otherwise.}
\end{cases}
\end{equation}
By contracting the copy tensor with its three outgoing interaction matrices, one obtains the local tensor \(T\) at each site:
\begin{equation}
    T_{ijklmn}=\sum_{a,b,c=1}^2\delta_{iakbmc}W_{aj}W_{bl}W_{cn}.
\end{equation}
By assembling these $T$ tensors on every lattice site, the original 3D Ising model is transformed into a 3D tensor network. 
The partition function $Z$ is then equivalent to the scalar value obtained by contracting all interconnected indices of this network.

\subsection{Approximate contraction with PEPS and PEPO}
\label{subsection:sm-tnmc-peps}

\begin{figure}[!t]
  \includegraphics[width=0.5\linewidth]{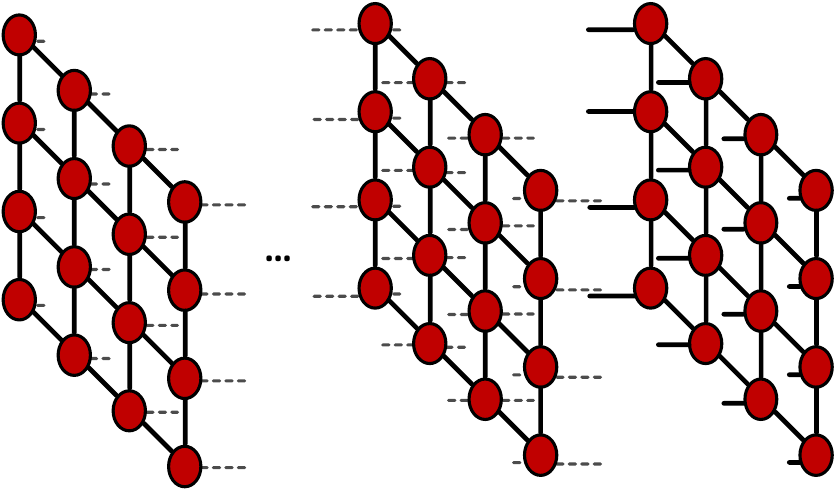}
  \caption{Projected Entangled Pair State (PEPS) and Operator (PEPO) representation of 3D systems.}
  \label{fig:3Dtensor}
\end{figure}

Directly contracting this 3D tensor network is computationally intractable for any non-trivial system size, as the cost scales exponentially. 
However, the structure of the network lends itself to efficient, approximate contraction schemes based on the concept of Projected Entangled Pair States (PEPS) and Operators (PEPO). 
Specifically, the 3D tensor network can be viewed as a stack of 2D layers, see Fig.~\ref{fig:3Dtensor}.
The partition function can be computed by sequentially contracting these layers from the top to the bottom layer.
This process can be formally expressed as applying a sequence of PEPOs, where each PEPO represents a 2D layer of the network, to a 2D PEPS that represents the accumulated contraction of the preceding layers.

\begin{figure}[!t]
  \includegraphics[width=0.5\linewidth]{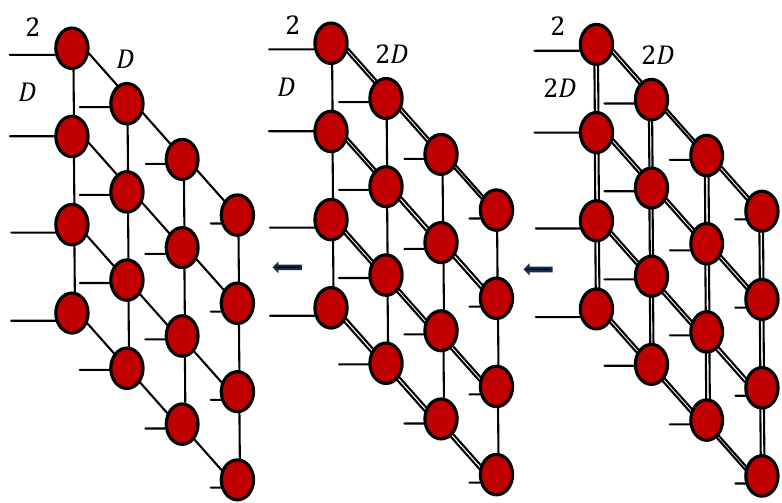}
  \caption{The contraction of a PEPO with a PEPS with bond dimension $D$ results in a new PEPS with bond dimension $2D$.
  Then the bond dimension is truncated back to $D$ via a singular value decomposition (SVD) applied across the horizontal and vertical indices.}
  \label{fig:PEPS}
\end{figure}

The contraction of a PEPO with a PEPS results in a new PEPS, but with a significantly larger virtual bond dimension, which is the dimension of the indices connecting the tensors. 
To maintain computational feasibility, this bond dimension must be truncated at each step. 
This is achieved through a controlled approximation based on the singular value decomposition (SVD). The tensor network along a bond is reshaped into a matrix, on which SVD is performed. 
This decomposition provides the optimal low-rank approximation of the matrix, and by retaining only the $D$ largest singular values and their associated singular vectors, the bond dimension is truncated to $D$. 
This procedure effectively compress the tensor network by integrating out the least significant degrees of freedom, with the accuracy of the entire calculation being systematically improvable by increasing $D$, see Fig.~\ref{fig:PEPS}.

After contracting all the layers in one of the spatial directions, the final result is a 2D tensor network, which can be readily contracted boundary MPS (bMPS) method, see~\cite{chen2025tensor} for a schematic illustration.
The accuracy and computational cost of the bMPS method are controlled by the maximum bond dimension of this boundary MPS, denoted by $\chi$.
At each step of the bMPS algorithm, a row of the MPO is applied to the boundary MPS, and the resulting MPS is then truncated back to the bond dimension $\chi$.
Although $D$ (PEPS bond dimension) and $\chi$ (boundary MPS bond dimension) are independent control parameters for the accuracy of the contraction, it is suggested that boundary MPS bond dimension should scale as $\chi \sim D^2$ to achieve reliable results.
If we consider this scaling, the leading computational cost is $\mathcal{O}(\chi^3 D^6) = \mathcal{O}(D^{12})$ that stems from the SVD in bMPS.

\subsection{Computation of conditional probabilities}
\label{subsection:sm-tnmc-cond}

The approximate tensor network contraction scheme is readily adapted for the computation of conditional probabilities, which are essential for the sampling process.
The conditional probability of a spin $s_i$ given the preceding spins in a predefined sequence, $\s_{<i}$, is expressed as the ratio of conditional partition functions:
\begin{equation}
    P_i(s_i|\s_{<i}) = \frac{\sum_{\s_{>i}}e^{-\beta E(\s)}}{\sum_{\s_{>i-1}}e^{-\beta E(\s)}} = \frac{Z(s_i,\s_{<i})}{\sum_{s_i}Z(s_i,\s_{<i})}.
\end{equation}
The conditional partition function, $Z(s_i,\s_{<i}) = \sum_{\s_{>i}}e^{-\beta E(\s)}$, is calculated using the same tensor network structure employed for the full partition function. 
However, the physical indices of the local tensors corresponding to the conditioned spins $\s_{<i}$ are fixed.
This is achieved by projecting the local tensor $T_j$ for each spin $s_j$ at a site $j<i$ onto a basis vector that represents the specific state of $s_j$. 
For instance, a spin state of $s_j=+1$ is represented by the vector $(1,0)^T$, while $s_j=-1$ corresponds to $(0,1)^T$.
Consequently, the conditional probability $P_i(s_i|\s_{<i})$ is determined by the ratio of two separate contractions of these modified tensor networks. 
The numerator is obtained from a network where the spins $\s_{<i}$ and $s_i$ are fixed to their specified values, while the denominator is the sum over the two possible states of $s_i$.

This procedure for computing conditional probabilities naturally defines an ancestral sampling protocol, which generates statistically independent samples from an approximate distribution $q(\s)$.
This distribution is factorized as a product of the conditional probabilities:
\begin{equation}
    q(\s) = \prod_{i=1}^{n} q(s_i|\s_{<i}),
    \label{eq:sm-proposal-distribution}
\end{equation}
forming the proposal distribution in a Markov chain.
At each step of this Markov chain, a candidate configuration $\s$ is generated independently of the current state $\s^{\prime}$. 
This candidate is then accepted according to the Metropolis-Hastings acceptance probability:
\begin{equation}
    A(\s^{\prime} \rightarrow \s) = \min \left\{1,  \frac{q(\s^{\prime}) \times e^{-\beta E(\s)}}{q(\s) \times e^{-\beta E(\s^{\prime})}}\right\}.
    \label{eq:sm-acceptance}
\end{equation}
The pseudocode for this 3D TNMC algorithm is provided in Algorithm~\ref{alg:tnmc3d}.
It is important to note that intermediate tensors generated during the contraction procedure can be cached and subsequently reused in the sampling process. 
Furthermore, due to the inherent independence of the samples, their generation can be efficiently parallelized.

\begin{algorithm}[!ht]
    \caption{Tensor Network Monte Carlo (TNMC) for 3D systems}
    \label{alg:tnmc3d}
    \SetKwInOut{Input}{Input}
    \SetKwInOut{Output}{Output}
    \SetKw{Procedure}{procedure}
    
    \Input{System size $L$, couplings $\{J_{ij}\}$, number of samples (Markov chain length) $N_s$, PEPS bond dimension $D$, boundary MPS bond dimension $\chi$, inverse temperature $\beta$}
    \Output{A Markov chain of spin configurations $\{\s^{(1)}, \dots, \s^{(N_s)}\}$}
    
    \BlankLine
    
    $\s^{(1)}, q(\s^{(1)}) \leftarrow \text{SampleConfiguration}(L, \{J_{ij}\}, D, \chi)$\;
    
    \BlankLine
    
    \For{$t \leftarrow 2$ \KwTo $N_s$}{
        $\s^{\prime}, q(\s^{\prime}) \leftarrow \text{SampleConfiguration}(L, \{J_{ij}\}, D, \chi)$\;
        
        $P_{\mathrm{acc}} \leftarrow \min \left\{1,  \frac{q(\s^{(t-1)}) \times e^{-\beta E(\s^{\prime})}}{q(\s^{\prime}) \times e^{-\beta E(\s^{(t-1)})}}\right\}$ \;
        
        \If{$\mathrm{random()} < P_{\mathrm{acc}}$}{
            $\s^{(t)} \leftarrow \s^{\prime}$\;
            $q(\s^{(t)}) \leftarrow q(\s^{\prime})$\;
        }
        \Else{
            $\s^{(t)} \leftarrow \s^{(t-1)}$\;
            $q(\s^{(t)}) \leftarrow q(\s^{(t-1)})$\;
        }
    }
    
    \BlankLine
    \hrule
    \BlankLine
    
    \SetKwProg{Fn}{procedure}{}{end}
    \Fn{$\mathrm{SampleConfiguration}(L, \{J_{ij}\}, D, \chi)$}{
        Initialize an empty spin configuration $\s=\{s_1,\cdots s_{L^3}\}$\;
        
        \For{$i \leftarrow 1$ \KwTo $L^3$}{
            Compute conditional probability $\hat{q} = q(s_i=+1 | \s_{<i})$ by contracting the 3D tensor network\;
            \If{$\mathrm{random()} < \hat{q}$}{
                $s_i \leftarrow +1$
            }
            \Else{
                $s_i \leftarrow -1$
            }
            $q(s_i | \s_{<i}) \leftarrow \hat{q}^{\delta_{s_i,+1}} (1-\hat{q})^{\delta_{s_i,-1}}$
        }
        
        $q(\s) \leftarrow \prod_i q(s_i | \s_{<i})$
        
        \KwRet{$\s, q(\s)$}\;
    }

\end{algorithm}

\begin{figure}[!t]
    \centering
    \includegraphics[width=0.5\linewidth]{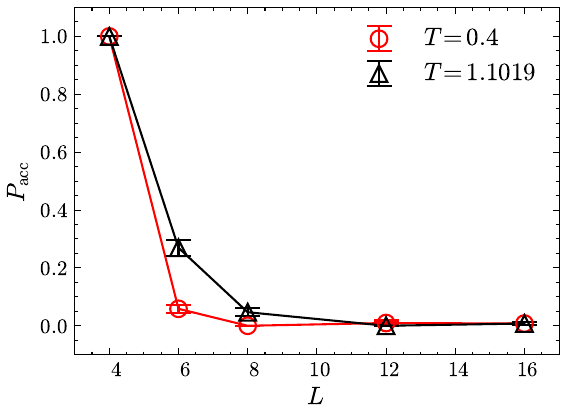}
    \caption{The Metropolis-Hasting acceptance probability $P_{\text{acc}}$ as a function of system size $L$ for the 3D EA model using the TNMC method ($\chi=4, D=4$).}
    \label{fig:Pacc-L}
\end{figure}

A significant drawback of the 3D TNMC algorithm, beyond its considerable computational and storage cost, is the precipitous decline in acceptance probability with increasing system size.
This is exemplified in the 3D EA model, where, as illustrated in Fig.~\ref{fig:Pacc-L}, the acceptance probability vanishes for system sizes as small as $L=12$ at the critical temperature $T_c = 1.1019$. 
The issue is exacerbated at lower temperatures, with the acceptance probability approaching zero for $L=8$ at $T=0.4$.
This behavior fundamentally restricts the applicability of the 3D TNMC algorithm for the simulation of large-scale 3D systems.

\section{Technical details of the Tensor Network Markov Chain Monte Carlo (TNMCMC)}
\label{section:sm-tnmcmc}

This section presents a detailed algorithmic exposition of the Tensor Network Markov Chain Monte Carlo (TNMCMC) method. 
We elaborate on the foundational principles of the method, including the partitioning of the spin lattice, the tensor network construction and contraction, and the specific implementations of two distinct update strategies, termed TNMCMC-1 and TNMCMC-2.

Instead of directly addressing the full 3D lattice at once, the system's degrees of freedom, $\s$, are partitioned into an ``active'' subset $\s_a$ to be sampled and a complementary ``held-fixed'' subset $\s_h$, such that $\s = \{\s_a, \s_h\}$.
The objective is then to sample a new active spins, $\s_a$, from the conditional probability of the Boltzmann distribution:
\begin{equation}
    P(\s_a|\s_h) = \prod_{i} P_i (\s_{a,i} | \s_{a,<i}, \s_h).
\end{equation}
The decomposition strategy leverages the concept of a Markov blanket, a fundamental principle from the field of probabilistic graphical models~\cite{bishop2006pattern}.
The Markov blanket of a given spin is the minimal set of neighboring spins that renders it conditionally independent of all other spins in the system.
Consequently, by fixing the state of the Markov blanket, the statistical properties of the enclosed spin are entirely determined, irrespective of the state of the wider system. 
The TNMCMC method extends this principle from a single spin to a multi-spin subsystem.

\begin{figure}[!t]
    \centering
    \includegraphics[width=0.5\linewidth]{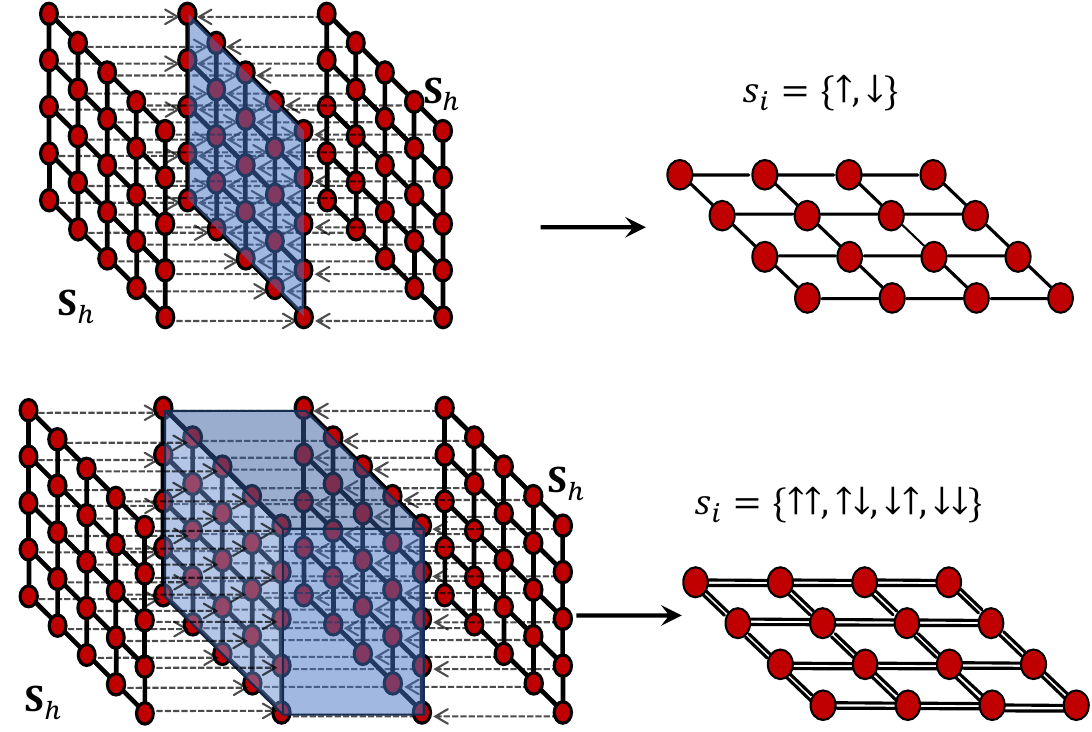}
    \caption{Illustration of TNMCMC-1 (top) and TNMCMC-2 (bottom).}
    \label{fig:tnmcmc-sampling}
\end{figure}

\subsection{TNMCMC-1: single-layer update strategy}
\label{subsection:sm-tnmcmc-1}

The first variant, TNMCMC-1, applies the Markov blanket concept to entire 2D layers within the 3D lattice. 
The 3D lattice is conceptualized as a vertical stack of 2D layers. 
For a given layer $L_i$, its spin configuration is conditionally independent of all non-adjacent layers $L_j$ (where $|i - j| > 1$) given the configurations of the immediately adjacent layers, $L_{i-1}$ and $L_{i+1}$. This pair, $\{L_{i-1}, L_{i+1}\}$, constitutes the Markov blanket for layer $L_i$.
By holding the spins in these Markov blanket layers fixed, their influence on layer $L_i$ is effectively reduced to a site-dependent external magnetic field (see Fig.~\ref{fig:tnmcmc-sampling}). 
The sampling problem for the spins in $L_i$ (the active set $\s_a$) is thereby transformed into an effective 2D statistical model, which can be efficiently sampled ancestrally using the bMPS method with the accuracy of this approximation being governed by the MPS bond dimension $\chi$, see Appendix~\ref{section:sm-tnmc} and Ref.~\cite{chen2025tensor} for details.
This process defines a non-local proposal distribution:
\begin{equation}
    q(\s_a | \s_h) = \prod_{i} q(\s_{a,i} | \s_{a,<i}, \s_h).
\end{equation}
To ensure convergence to the true Boltzmann distribution, the proposed layer update from $\s_a'$ to $\s_a$ (with $\s_h$ fixed) is accepted or rejected according to the Metropolis-Hastings criterion:
\begin{align}
    &A\left(\{\s_a^\prime,\s_h\} \rightarrow \{\s_a,\s_h\}\right) \\=& \min \left\{
    1, \frac{q(\s_a^\prime|\s_h) \times e^{-\beta E(\{\s_a,\s_h\})}}{q(\s_a | \s_h) \times e^{-\beta E(\{\s_a^\prime,\s_h\})}}
    \right\}.
\end{align}
A complete Monte Carlo step involves the sequential application of this update procedure to every layer in the system.
To guarantee ergodicity and an isotropic exploration of the state space, this process is iterated, with updates cycling through the $x$, $y$, and $z$ axes of the lattice.
The pseudocode is provided in Algorithm~\ref{alg:tnmcmc-1}.
Note that the TNMCMC framework is also inherently amenable to parallel implementation. 
For example, in TNMCMC-1 one may alternately fix all spin variables on the odd or even layers and update the complementary set in parallel, see Sec.~\ref{subsection:sm-results-parallel} for numerical results.

\begin{algorithm}[!ht]
    \caption{Tensor Network Markov Chain Monte Carlo (TNMCMC-1)}
    \label{alg:tnmcmc-1}
    \SetKwInOut{Input}{Input}
    \SetKwInOut{Output}{Output}
    \SetKw{Procedure}{procedure}
    
    \Input{System size $L$, couplings $\{J_{ij}\}$, Markov chain length $N_s$, MPS bond dimension $\chi$, inverse temperature $\beta$}
    \Output{A Markov chain of spin configurations $\{\s^{(1)}, \dots, \s^{(N_s)}\}$}
    
    \BlankLine
    
    Initialize spin configuration $\s^{(0)}$
    
    \BlankLine
    
    \For{$t \leftarrow 1$ \KwTo $N_s$}{
        Select a sampling direction \{$x,y,z$\}\;
        \For{$l \leftarrow 1$ \KwTo $L$}{
            Let $\s_a$ be the spin configuration of the $l$-th layer\; 
            Let $\s_h$ be the spin configuration of the $(l-1)$-th and $(l+1)$-th layer\; 
            $\s_a^{\prime}, q(\s_a^{\prime}|\s_h), q(\s_a|\s_h) \leftarrow \text{SampleLayer}(L, \{J_{ij}\}, \chi, \s_a, \s_h )$\;
            $P_{\mathrm{acc}} = \min \left\{1, \frac{q(\s_a^\prime|\s_h) \times e^{-\beta E(\{\s_a,\s_h\})}}{q(\s_a | \s_h) \times e^{-\beta E(\{\s_a^\prime,\s_h\})}}\right\}$\;
            
            \If{$\mathrm{random()} < P_{\mathrm{acc}}$}{
            $\s_a^{(t)} \leftarrow \s_a^{\prime}$
            }
            \Else{
                $\s_a^{(t)} \leftarrow \s_a^{(t)}$
            }
        }
        Record the configuration $\s^{(t)}$
    }
    
    \BlankLine
    \hrule
    \BlankLine
    
    \SetKwProg{Fn}{procedure}{}{end}
    \Fn{$\mathrm{SampleLayer}(L, \{J_{ij}\}, \chi,\s_a,\s_h)$}{
        Construct the 2D tensor network given $\s_h$ and $\{J_{ij}\}$\; 
        Initialize an empty spin configuration $\s = \{s_1, \cdots,s_{L^2}\}$\;
        
        \For{$i \leftarrow 1$ \KwTo $L^2$}{
            Compute conditional probability $q(s_i=+1 | \s_{<i})$ by contracting the 2D tensor network\;
            Sample $s_i$ from $q(s_i=+1 | \s_{<i})$\;
            Append $s_i$ to $\s$\;
        }
        Evaluate proposal probabilities for $q(\s)$ and $q(\s_a)$\;
        \KwRet{$\s, q(\s), q(\s_a)$}\;
    }

\end{algorithm}

\subsection{TNMCMC-2: coarse-graining with superspins}
\label{subsection:sm-tnmcmc-2}

The second strategy, TNMCMC-2, utilizes a coarse-graining procedure to more accurately capture inter-layer correlations. 
In this approach, two adjacent layers, for instance, $L_i$ and $L_{i+1}$, are grouped and treated as a single, effective composite layer. 
For an Ising system with binary spin states, each ``superspin'' in this composite layer possesses a local state space of dimension $2 \times 2 = 4$ (see Fig.~\ref{fig:tnmcmc-sampling}).

This coarse-graining technique enables the TNMCMC-2 algorithm to manage the strong inter-layer couplings between the grouped layers within the 2D tensor network framework. 
When sampling the composite layer, the intra-layer couplings of both $L_i$ and $L_{i+1}$, as well as the inter-layer couplings that connect them, are all absorbed into the local tensor definitions of the new 2D tensor network. 
The Markov blanket for this composite layer is now constituted by layers $L_{i-1}$ and $L_{i+2}$.
The sampling procedure is analogous to TNMCMC-1, see Algorithm~\ref{alg:tnmcmc-1}: the spins of the Markov blanket are fixed, and the resulting 2D TN of superspins are sampled sequentially, followed by a Metropolis-Hastings acceptance step. 
Although this approach is more computationally demanding, due to the larger local state space of the superspins necessitating a significantly larger MPS bond dimension, $\chi$, to achieve comparable approximation accuracy, it provides a more potent update mechanism. 
By directly incorporating the correlations of the third dimension into the sampling proposal, TNMCMC-2 offers a more powerful tool for probing the complex physics of 3D systems.

\subsection{Comparison between TNMC-3D and TNMCMC}
\label{subsection:sm-tnmcmc-comparison}

Here we discuss the main distinctions between the TNMC-3D approach and our proposed TNMCMC algorithm.

\textbf{Computational complexity.} The leading computational cost of TNMC-3D scales as $\mathcal{O}(\chi^3 D^6)$, where $\chi$ is the boundary MPS bond dimension and $D$ is the PEPS bond dimension.
For typical scenarios where $\chi \sim D^2$, this complexity scales unfavorably as $\mathcal{O}(D^{12})$.
In stark contrast, a full sweep over the entire lattice, which constitutes a single Monte Carlo step encompassing all $L^3$ sites, exhibits a much more favorable scaling of $\mathcal{O}(\chi^3 L^3)$ for TNMCMC.

\textbf{Sampling scheme and statistical independence.} The TNMC-3D method generates a set of statistically independent samples by repeatedly contracting the full 3D tensor network. 
This process implicitly defines a Markov chain where each state is an entirely new, uncorrelated configuration of the system.
Conversely, the TNMCMC algorithm constructs an explicit Markov chain by performing local updates on one or two layers of spins while the remainder of the lattice is held fixed. 
This sequential updating scheme introduces temporal correlations between successive configurations, meaning that the samples generated by TNMCMC are not statistically independent in the same manner as those produced by TNMC-3D. 

\begin{figure}[!t]
    \centering
    \includegraphics[width=0.5\linewidth]{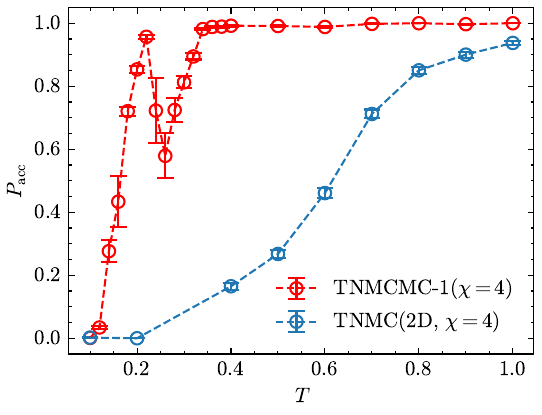}
    \caption{The Metropolis-Hasting acceptance probability $P_{\mathrm{acc}}$ as a function of temperature $T$ for the 3D EA model using TNMCMC-1 with bond dimension $\chi=4$.
    For comparison, results for the 2D EA model obtained via TNMC with bond dimension $\chi=4$ are also shown.
    All simulations were conducted on systems of linear size $L = 32$.}
    \label{fig:Pacc-T}
\end{figure}

\textbf{Low-temperature efficiency.}  The practical advantage of the TNMCMC algorithm becomes particularly pronounced at low temperatures. 
To demonstrate this, we performed simulations of the 3D EA spin-glass model utilizing the TNMCMC-1 method. 
For comparative purposes, we also applied the conventional TNMC method to the 2D EA model. 
As depicted in Fig.~\ref{fig:Pacc-T}, the Metropolis-Hastings acceptance probability for the TNMC method experiences a rapid decline as the temperature is lowered.
In contrast, the TNMCMC-1 algorithm maintains a significantly higher acceptance probability even at considerably lower temperatures, highlighting its superior efficiency in exploring the low-temperature regime of complex spin systems.

\section{Additional results for 3D statistical models}
\label{section:sm-results}
This section presents supplementary results that further underscore the efficacy and versatility of the TNMCMC algorithm. 
We provide a more detailed analysis of its performance on the 3D Ising model, the Edwards-Anderson (EA) model, and the 3-state Potts model, thereby demonstrating its advantages across a range of statistical systems. 
To provide a quantitative assessment of the sampling efficiency, we employ two key metrics whose rigorous definitions are provided below.

\textbf{Integrated autocorrelation time.} In the context of MCMC simulations, the integrated autocorrelation time, denoted as $\tau_{\text{int},O}$ for a specific observable $O$, is a crucial measure of the efficiency of the sampling process. 
It quantifies the number of simulation steps required for the configurations generated by the Markov chain to become effectively uncorrelated. 
The autocorrelation function $\rho_{O}(k)$ of the observable $O(t)$ is computed as:
\begin{equation}
    \rho_O(k) = \frac{\langle O(t) O(t+k) \rangle - \langle O(t) \rangle^2}{\langle O^2(t) \rangle - \langle O(t) \rangle^2},
\end{equation}
where $t$ and $k$ are measured in MC steps.
The integrated autocorrelation time $\tau_{\text{int},O}$ is then formally defined as the summation of the autocorrelation function over all non-negative lag times:
\begin{equation}
    \tau_{\text{int},O} = \frac{1}{2} + \sum_{k=1}^{\infty} \rho_O(k).
\end{equation}
In practice, this summation is truncated at a cutoff window where $\rho_O(k)$ ceases to be statistically significant.

\textbf{First-passage time.} The second metric, the first-passage time $\tau_{\text{fp},O}$, quantifies the Monte Carlo steps required for an observable $O$ to first reach a specific target value, $O_{\text{target}}$, from a defined initial state $O(t=0)$. 
It is formally defined as:
\begin{equation}
    \tau_{\text{fp},O} = \min\left\{ t>0 \vert O(t) = O_{\text{target}}\right\}.
\end{equation}
This metric is particularly useful for characterizing the time required to escape from metastable states within the system's energy landscape. 
In our experiments, the direct computation of the integrated autocorrelation time for the overlap parameter, $\tau_{\text{int},q}$, is computationally prohibitive for large systems. 
Therefore, we utilize the first-passage time to zero magnetization $\tau_{\text{fp},M}$ as a proxy. 
Specifically, we initialize the simulation in a fully magnetized state (magnetization $M=1$) and measure the number of Monte Carlo steps required to reach a state of zero magnetization for the first time.

\subsection{The 3D Ising model}
\label{subsection:sm-results-ising}

\begin{figure}[!t]
    \centering
    \includegraphics[width=0.5\linewidth]{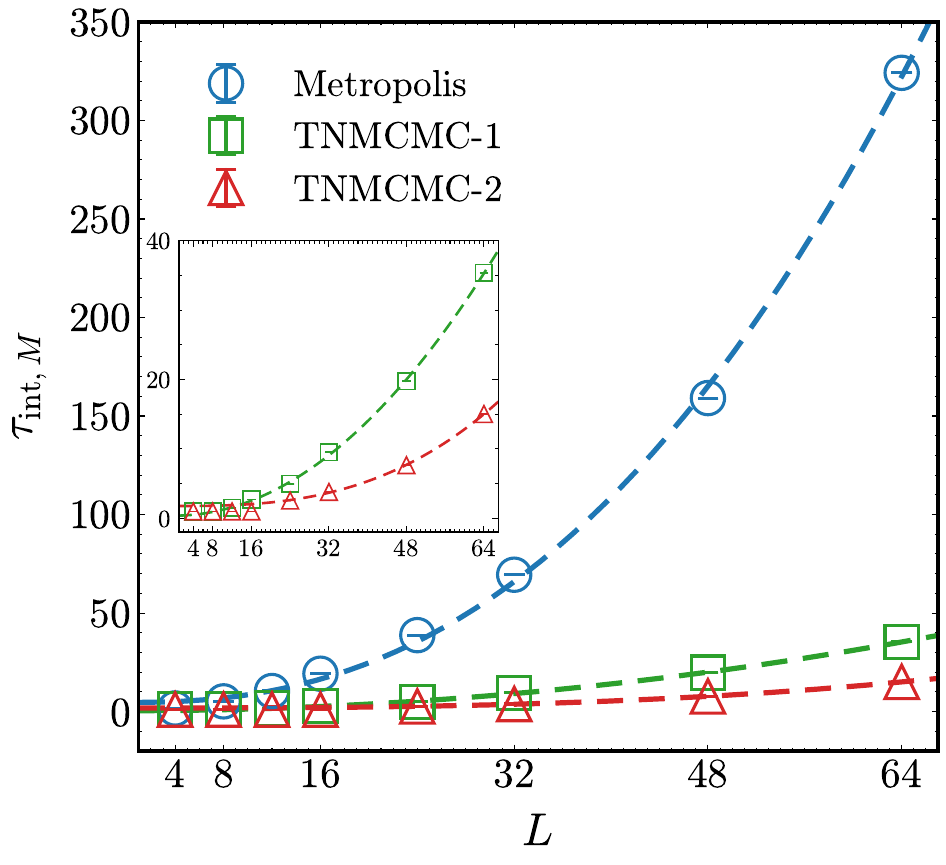}
    \caption{Integrated autocorrelation time of the magnetization, $\tau_{\text{int},M}$, as a function of the system size $L$ at its critical temperature $T_c$ for the 3D Ising model.
    The performance of the TNMCMC algorithm, with bond dimensions $\chi=2$ (TNMCMC-1, green) and $\chi=4$ (TNMCMC-2, red) is compared with the standard Metropolis algorithm (blue). 
    }
    \label{fig:3DIsingTc_tau}
\end{figure}

To evaluate the broader applicability of the TNMCMC algorithm beyond the complex energy landscapes of spin glasses and systems with first-order phase transitions, we have applied it to the 3D Ising model.
This model is a canonical example of a system exhibiting a continuous, or second-order, phase transition. 
Such transitions are characterized by a correlation length that diverges at the critical temperature, $T_c$, leading to a phenomenon known as critical slowing down in simulations.
The main challenge in simulating systems near a phase transition is the dramatic increase in the autocorrelation time, which signifies that local update algorithms, such as the Metropolis method, require a prohibitively large number of steps to generate statistically independent configurations.

In Fig.~\ref{fig:3DIsingTc_tau}, our results for the integrated autocorrelation time of the magnetization, $\tau_{\text{int},M}$, as a function of the system size $L$ at $T_c$ confirm the expected power-law scaling for the Metropolis algorithm, where $\tau_{\text{int},M} \sim L^z$. 
In stark contrast, both the TNMCMC-1 and TNMCMC-2 algorithms demonstrate a significant reduction in $\tau_{\text{int},M}$. 
The TNMCMC-2 algorithm, which employs more extensive, collective updates, is particularly effective in suppressing this critical slowing down.
The efficiency of TNMCMC stems from its ability to propose global updates by leveraging a tensor network representation of the system, which can capture the long-range correlations present at criticality. 

It should be noted that cluster algorithms such as Swendsen-Wang~\cite{swendsen1987nonuniversal} or Wolff~\cite{wolff1989collective} are known to eliminate critical slowing down in the 3D Ising model by flipping large clusters of correlated spins, but they rely heavily on specific symmetries and spin interactions. 
The TNMCMC framework, on the other hand, provides a general-purpose and powerful alternative. It does not rely on problem-specific symmetries, offering a versatile non-local update scheme.

\subsection{The 3D Edwards-Anderson (EA) model}
\label{subsection:sm-results-ea}

\begin{figure*}[!t]
    \centering
    \includegraphics[width=\linewidth]{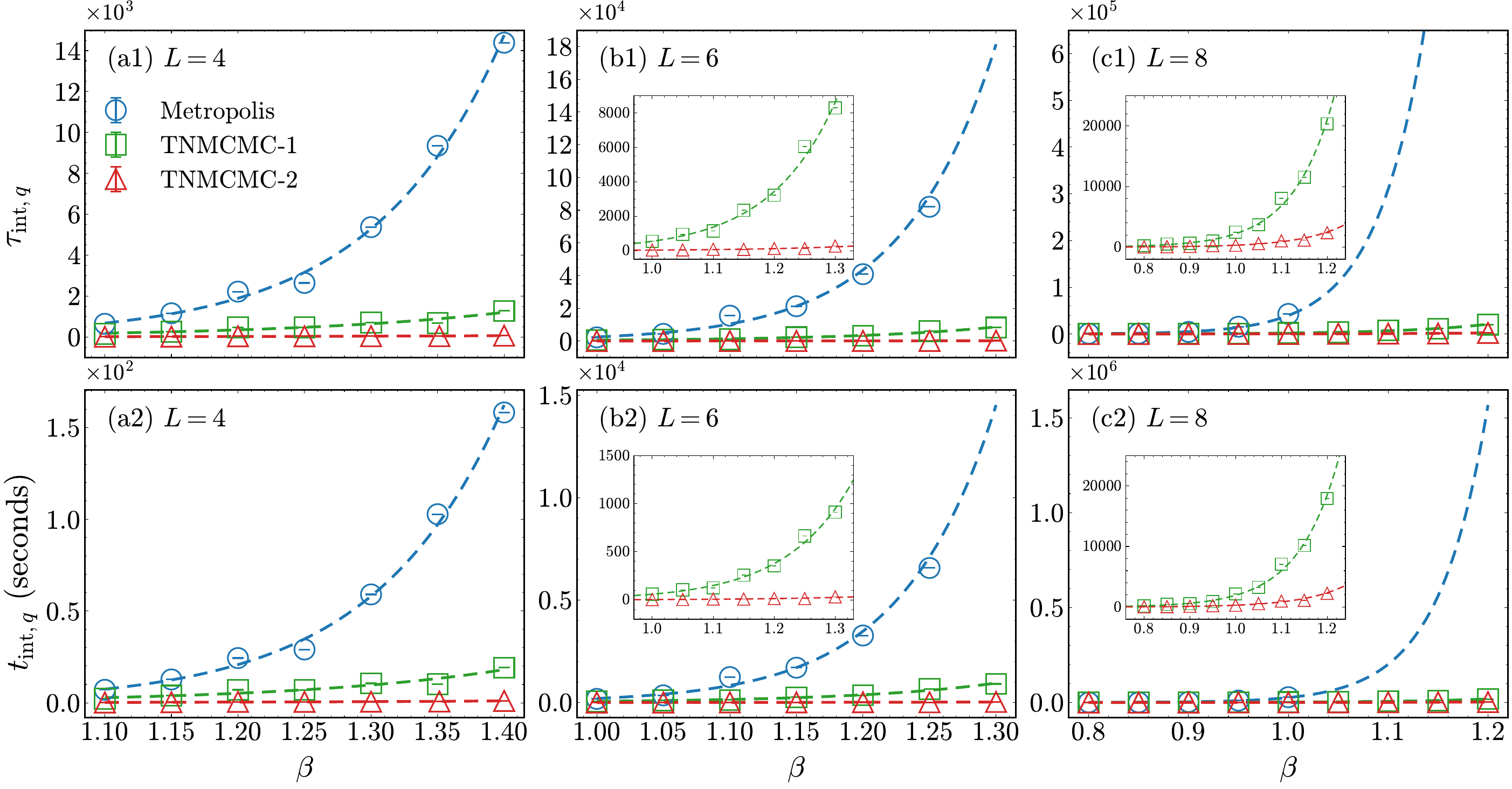}
    \caption{Top row: Integrated autocorrelation time, $\tau_{\text{int},q}$, for the spin-glass overlap parameter. Bottom row: Corresponding computational cost, $t_{\text{int},q} = \tau_{\text{int},q} \times t_s$, where $t_s$ is the wall-clock time per sweep. 
    Data are presented for the 3D EA model with system sizes $L=4$, 6, and 8 across a range of temperatures.}
    \label{fig:tauT}
\end{figure*}

\textbf{Autocorrelation versus temperature.} A quantitative benchmark of the TNMCMC algorithm againt the conventional Metropolis method is performed by investigating the temperature and system-size dependence of the simulation dynamics.
We first examine the integrated autocorrelation time, $\tau_{\text{int},q}$, of the spin-glass overlap parameter as a function of temperature for system sizes $L=4, 6,$ and $8$, as depicted in Fig.~\ref{fig:tauT}.
The corresponding computational cost to generate a statistically independent sample, defined as the CPU time $t_{\text{int},q} = \tau_{\text{int},q} \times t_s$, is also presented. 

For the conventional Metropolis algorithm, a pronounced increase in the integrated autocorrelation time is observed as the temperature is lowered.
This behavior is characteristic of the critical slowing down encountered in the spin-glass phase, where the rugged energy landscape and the local nature of Metropolis updates impede efficient exploration of the state space. 
Consequently, simulations using the Metropolis algorithm for $L=8$ were computationally prohibitive for inverse temperatures $\beta > 1.0$. In contrast, the TNMCMC algorithm, which leverages a tensor-network-based global update scheme, effectively circumvents large energy barriers, leading to a more efficient sampling of metastable states.

\begin{figure}[!t]
    \centering
    \includegraphics[width=1\linewidth]{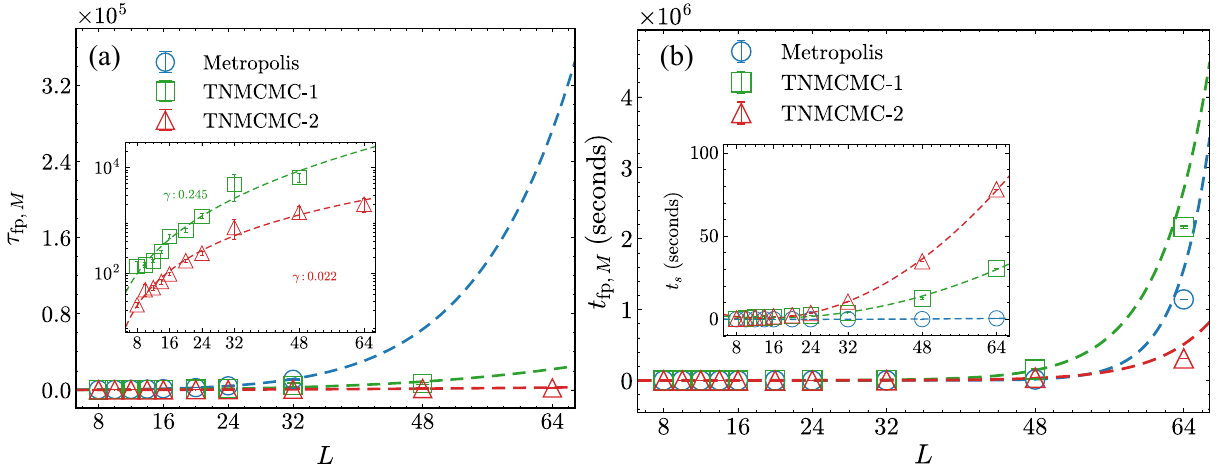}
    \caption{Algorithmic performance comparison for the 3D EA model at an inverse temperature of $\beta=1.0$. (a) The first-passage time to zero magnetization, $\tau_{\text{fp}, M}$, is plotted against system size $L$. The inset displays the scaling for the TNMCMC method, which is consistent with the functional form $\tau_{\text{fp},M} \approx \tau_0 \exp(cL^\psi)$, characteristic of activated dynamics in a glassy phase. (b) A comparison of the computational cost $t_{\text{fp},M} = \tau_{\text{fp},M} \times t_s$. The inset shows the wall-clock time $t_s$ required for a single sweep for each algorithm.}
    \label{fig:tauL}
\end{figure}

\textbf{Autocorrelation versus system size.} A more detailed analysis of the autocorrelation properties is conducted at a fixed inverse temperature $\beta=1.0$, as shown in Fig.~\ref{fig:tauL}.
The direct calculation of $\tau_{\text{int},q}$ becomes intractable for larger system sizes. Therefore, to characterize the decorrelation time, we utilize the first-passage time to zero magnetization, $\tau_{\text{fp},M}$.
The scaling of $\tau_{\text{fp},M}$ with system size offers crucial insights into the nature of the underlying dynamics. 
In glassy systems, this relationship is often characterized by an exponential scaling, indicative of an activated dynamics where the system must overcome energy barriers. 
This behavior is well-described by the following functional form:
\begin{equation}
  \tau_{\text{fp},M} \approx \tau_0 \exp(c L^\psi),
\end{equation}
where $\psi$ is the barrier exponent.
The exponential dependence on a power of the system size highlights a much more dramatic slowing down than the power-law scaling ($\tau \sim L^z$) observed in conventional critical slowing down systems.
A fit of our numerical data to this equation yields distinct scaling exponents for the different algorithms. 
For the conventional Metropolis algorithm, we find $\psi \approx 0.57$. 
In contrast, the TNMCMC-1 and TNMCMC-2 methods exhibit significantly smaller exponents of $\psi \approx 0.245$ and $\psi \approx 0.022$, respectively. 
This substantial reduction in the scaling exponent underscores the superior efficiency of the TNMCMC approach in navigating the complex energy landscape of the 3D EA spin glass.

The inset of Fig.~\ref{fig:tauL} displays the average wall-clock time per Monte Carlo step, $t_s$, for each of the three algorithms. 
It is observed that $t_s$ for the TNMCMC methods increases with system size. 
This is an inherent consequence of the computational complexity of the TNMCMC algorithm itself.
Nevertheless, the significant reduction in the autocorrelation time overwhelmingly compensates for this increase in single-step computational cost.
Consequently, when considering the total computational time required to generate statistically independent configurations, the TNMCMC methods demonstrate a substantial advantage over the traditional Metropolis algorithm, particularly for larger system sizes.

\subsection{The 3D 3-state Potts model}
\label{subsection:sm-results-potts}

\begin{figure}[!t]
    \centering
    \includegraphics[width=0.5\linewidth]{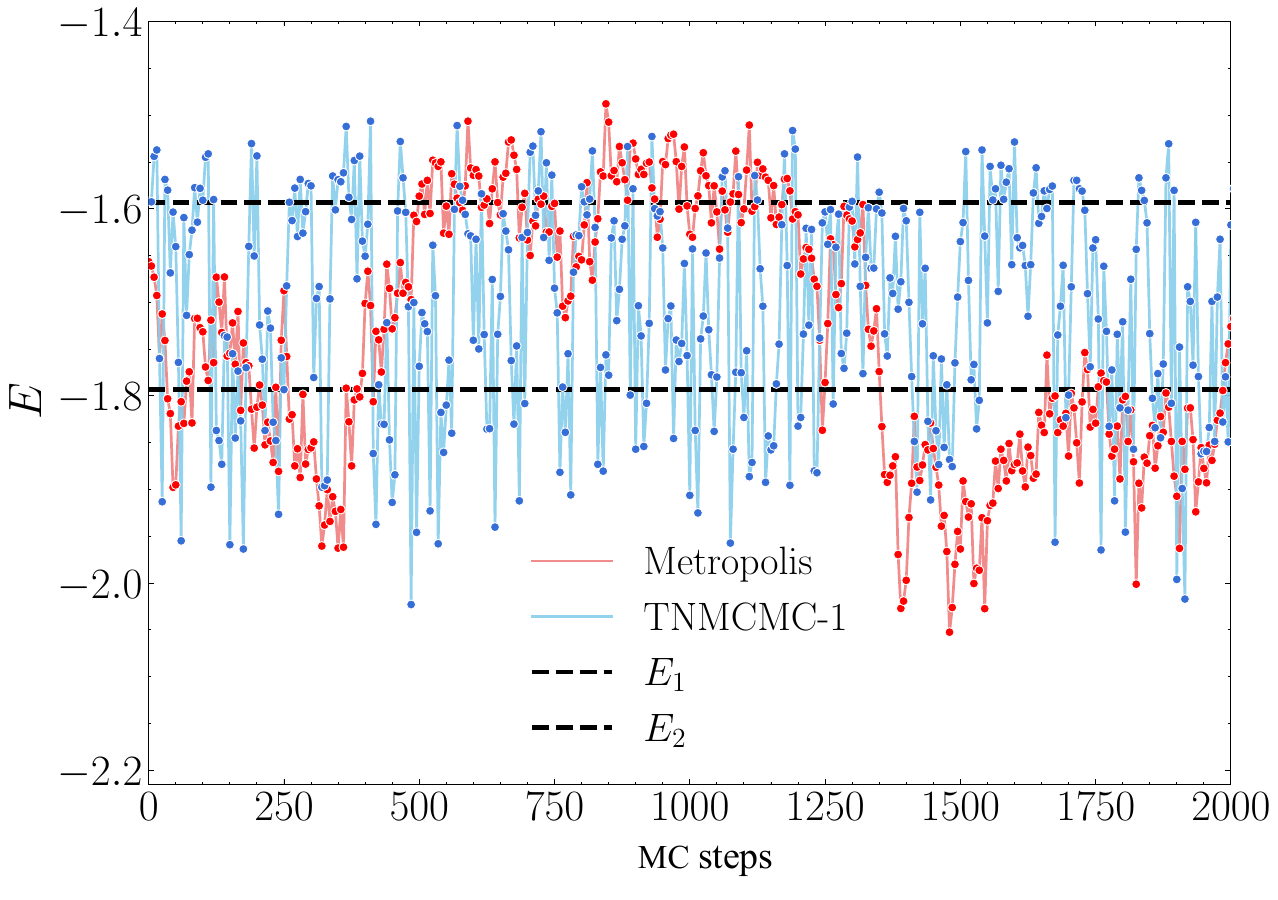}
    \caption{Energy time series obtained via Metropolis (red) and TNMCMC-1 (blue) simulations at $T_c$ for system size $L=16$.}
    \label{fig:energy_evolution}
\end{figure}

In Fig.~\ref{fig:energy_evolution}, we demonstrate the superior sampling efficiency of our TNMCMC algorithm by examining the equilibrium dynamics for a 3D 3-state Potts model with $L=16$.
Even with a minimal bond dimension of $\chi=3$, our TNMCMC-1 algorithm demonstrates remarkable performance. 
The system's energy fluctuates frequently between the two well-defined levels that signify the coexisting ordered and disordered phases, indicating that the algorithm effectively navigates the phase space.
Conversely, the conventional Metropolis algorithm, which relies on local updates, displays strong trapping behavior. 
The simulation remains confined to a single thermodynamic phase for extended periods, on the order of $10^3$ Monte Carlo steps
The results clearly establish the crucial role of non-local updates, as provided by TNMCMC, in overcoming the free-energy barriers that frustrate conventional sampling methods.

\subsection{Parallelism of TNMCMC}
\label{subsection:sm-results-parallel}

\begin{figure}[!t]
    \centering
    \includegraphics[width=1\linewidth]{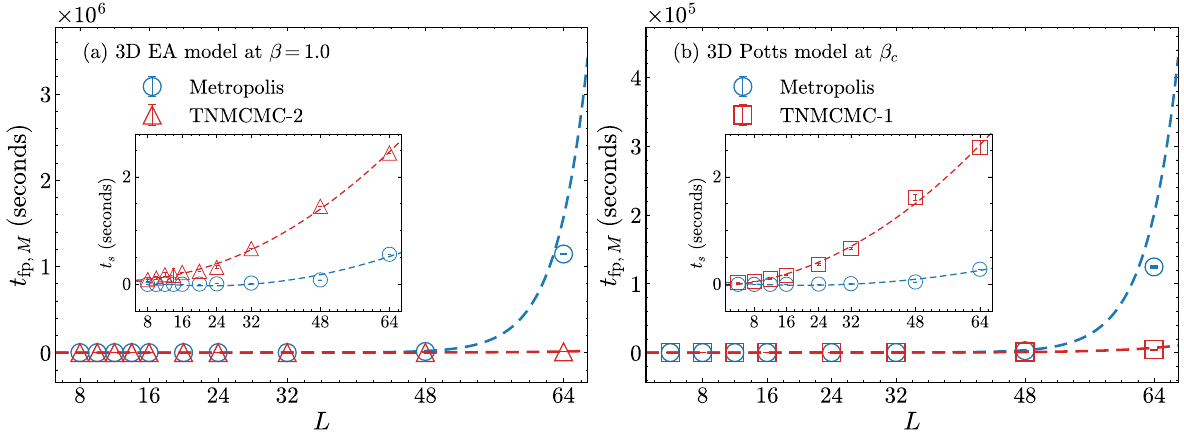}
    \caption{Idealized parallel performance of the TNMCMC algorithm. The figure displays the first-passage time to zero magnetization, $t_{\text{fp},M}$, and the single-sweep wall-clock time, $t_s$, for the 3D EA and 3-state Potts models, assuming perfect parallel execution across independent layers.}
    \label{fig:parallelism}
\end{figure}

As discussed in Sec.~\ref{subsection:sm-tnmcmc-1}, an important feature of the TNMCMC framework is its inherent parallelism. 
The structure of the 3D lattice, governed by the Markov blanket property, permits the independent and simultaneous updating of distinct layers. 
This characteristic allows for a layer-level parallel implementation, which can substantially reduce the simulation time.
To quantify the benefits of this parallelism, we assess the idealized parallel performance of the TNMCMC method. 
This is achieved by estimating the wall-clock time required for a single, complete sweep update, under the assumption of perfect parallel execution across all independent layers. 
Figure~\ref{fig:parallelism} presents the estimated computational cost for both the 3D EA and 3-state Potts models.
The inset of Fig.~\ref{fig:parallelism}(a) details the average time, $t_s$, for a single sweep update. 
A noteworthy trend emerges: the sweep time ratio, $t_s(\text{TNMCMC-2})/t_s(\text{Metropolis})$, exhibits a pronounced decrease with increasing system size, plummeting from 354.5 at $L = 8$ to a mere 9.4 at $L = 64$. 
A similar behavior is observed for the TNMCMC-1 method, where the corresponding ratio, $t_s(\text{TNMCMC-1})/t_s(\text{Metropolis})$, declines from 235 at $L = 8$ to only 9 at $L = 64$. 
This demonstrates that the parallel execution effectively amortizes the increased computational complexity inherent in the tensor-network-based updates.

These results affirm that while the serial implementation of TNMCMC methods entails a higher computational cost per update, the algorithm's architecture is exceptionally well-suited for parallelization. 
The relative performance of TNMCMC, in terms of both per-update time and total equilibration time, improves markedly with growing system size. 
This scalability renders TNMCMC not only a theoretically powerful sampling methodology but also a highly practical numerical tool for contemporary parallel computing platforms. 
Future implementations on GPUs or distributed computing systems are anticipated to further enhance these performance gains.

\bibliography{references.bib}